\documentclass[reqno,openright,dvips,11pt,myheadings,twoside]{amsart}

\usepackage{ifthen}

\newcounter{dtlForSubmission} 
\newcounter{dtlMarginComments} 
\newcounter{dtlSomeDetail} 
\newcounter{dtlFullDetails} 

\newcounter{DetailLevel} 

\newcommand{\DetailMarginNote}[1]{
    \ifthenelse{\value{DetailLevel}=\value{dtlMarginComments} \or \value{DetailLevel}>\value{dtlMarginComments}}
        {{\small #1}}{}
    }

\newcommand{\DetailSome}[1]{
    \ifthenelse{\value{DetailLevel}=\value{dtlSomeDetail} \or \value{DetailLevel}>\value{dtlSomeDetail}}
        {{\small \textbf{Detailed compile only}: #1}}{}
    }

\newcommand{\DetailFull}[1]{
    \ifthenelse{\value{DetailLevel}=\value{dtlFullDetails} \or \value{DetailLevel}>\value{dtlFullDetails}}
        {{\small \textbf{Detailed compile only}: #1}}{}
    }

\newcommand{\NotDetailSome}[1]{
    \ifthenelse{\value{DetailLevel}=\value{dtlSomeDetail} \or \value{DetailLevel}>\value{dtlSomeDetail}}
        {}{#1}
    }

\newcommand{\NotDetailFull}[1]{
    \ifthenelse{\value{DetailLevel}=\value{dtlFullDetails} \or \value{DetailLevel}>\value{dtlFullDetails}}
        {}{#1}
    }

\newcommand{\DetailSomeElse}[2]{
    \ifthenelse{\value{DetailLevel}=\value{dtlSomeDetail} \or \value{DetailLevel}>\value{dtlSomeDetail}}
        {{\small \textbf{Detailed compile only}: #1}}{#2}
    }

\newcommand{\DetailFullElse}[2]{
    \ifthenelse{\value{DetailLevel}=\value{dtlFullDetails} \or \value{DetailLevel}>\value{dtlFullDetails}}
        {{\small \textbf{Detailed compile only}: #1}}{#2}
    }

\newcommand{\DetailSomeInline}[1]{
    \ifthenelse{\value{DetailLevel}=\value{dtlSomeDetail} \or \value{DetailLevel}>\value{dtlSomeDetail}}
        {{\small #1}}{}
    }

\newcommand{\DetailFullInline}[1]{
    \ifthenelse{\value{DetailLevel}=\value{dtlFullDetails} \or \value{DetailLevel}>\value{dtlFullDetails}}
        {{\small #1}}{}
    }

\newcommand{\DetailSomeElseInline}[2]{
    \ifthenelse{\value{DetailLevel}=\value{dtlSomeDetail} \or \value{DetailLevel}>\value{dtlSomeDetail}}
        {{\small #1}}{#2}
    }

\newcommand{\DetailFullElseInline}[2]{
    \ifthenelse{\value{DetailLevel}=\value{dtlFullDetails} \or \value{DetailLevel}>\value{dtlFullDetails}}
        {{\small #1}}{#2}
    }

\newcommand{\ExplainDetailLevel}{
    Detail level is
    \ifthenelse{\value{DetailLevel}=\value{dtlForSubmission}}
        {0: for submission}
        {\ifthenelse{\value{DetailLevel}=\value{dtlMarginComments}}
            {1: as for submission but with margin comments}
            {\ifthenelse{\value{DetailLevel}=\value{dtlSomeDetail}}
                {2: some proofs not intended for submission}
               {\ifthenelse{\value{DetailLevel}=\value{dtlFullDetails}}
                   {3: full details}
                   {invalid}
                }
            }
        }
    }

\newtheorem{theorem}{Theorem}[section]
\newtheorem{lemma}[theorem]{Lemma}
\newtheorem{cor}[theorem]{Corollary}

\theoremstyle{definition}
\newtheorem{definition}[theorem]{Definition}

\newtheorem{remark}{Remark}[section]

\numberwithin{equation}{section}

\newcommand{\abs}[1]{\left\vert#1\right\vert}

\newcommand{\innp}[1]{\ensuremath{\left< #1 \right>}}

\newcommand{\BoldTau}
    {\mbox{\boldmath $\tau$}}

\newcommand{\skipline}{\vspace{11pt}}

\newcommand{\BB}[1]{\ensuremath{\mathbb{#1}}}
\newcommand{\R}{\ensuremath{\BB{R}}} %
\newcommand{\Z}{\ensuremath{\BB{Z}}} %
\newcommand{\iny}{\ensuremath{\infty}}
\newcommand{\grad}{\ensuremath{\nabla}}
\DeclareMathOperator{\dv}{div} %
\DeclareMathOperator{\curl}{curl} %
\newcommand{\wh}{\ensuremath{\widehat}}
\newcommand{\prt}{\ensuremath{\partial}}
\newcommand{\brac}[1]{\ensuremath{\left[ #1 \right]}}
\newcommand{\pr}[1]{\ensuremath{\left( #1 \right) }}
\newcommand{\set}[1]{\ensuremath{\left\{ #1 \right\}}}
\newcommand{\smallset}[1]{\ensuremath{\{ #1 \}}}
\newcommand{\norm}[1]{\ensuremath{\left\Vert #1 \right\Vert}}
\newcommand{\smallnorm}[1]{\ensuremath{\Vert #1 \Vert}}
\newcommand{\refS}[1]{Section~\ref{S:#1}}
\newcommand{\refT}[1]{Theorem~\ref{T:#1}}
\newcommand{\refL}[1]{Lemma~\ref{L:#1}}
\newcommand{\refD}[1]{Definition~\ref{D:#1}}
\newcommand{\refC}[1]{Corollary~\ref{C:#1}}
\newcommand{\refE}[1]{Equation~(\ref{e:#1})}
\newcommand{\refR}[1]{Remark~(\ref{R:#1})}
\newcommand{\eps}{\ensuremath{\epsilon}}
\newcommand{\Cal}[1]{\ensuremath{\mathcal{#1}}}
\newcommand{\al}{\ensuremath{\alpha}}
\newcommand{\la}{\ensuremath{\lambda}}
\newcommand{\pdx}[2]{\frac{\prt #1}{\prt #2}}
\newcommand{\pdxtwo}[2]{\frac{\prt^2 #1}{\prt {#2}^2}}
\newcommand{\diff}[2]{\frac{ d#1}{d#2}}
\newcommand{\difftwo}[2]{\frac{d^2#1}{d{#2}^2}}
\newcommand{\ol}{\overline}
\newcommand{\smallabs}[1]{\ensuremath{\vert #1 \vert}}

\begin{document}

\raggedbottom

\numberwithin{equation}{section}

%
%
\newcommand{\MarginNote}[1]{
    \marginpar{
        \begin{flushleft}
            \footnotesize #1
        \end{flushleft}
        }
    }

%
%
\newcommand{\NoteToSelf}[1]{
    }

%
%
\newcommand{\Obsolete}[1]{
    }

%
%
\newcommand{\Tentative}[1]{
    }

\newcommand{\Detail}[1]{
    \MarginNote{Detail}
    \skipline
    \hspace{+0.25in}\fbox{\parbox{4.25in}{\small #1}}
    \skipline
    }

\newcommand{\Todo}[1]{
    \skipline \noindent \textbf{TODO}:
    #1
    \skipline
    }

\newcommand{\Comment}[1] {
    \skipline
    \hspace{+0.25in}\fbox{\parbox{4.25in}{\small \textbf{Comment}: #1}}
    \skipline
    }

%
%

\newcommand{\IntTR}
    {\int_{t_0}^{t_1} \int_{\R^d}}

\newcommand{\IntAll}
    {\int_{-\iny}^\iny}

\newcommand{\Schwartz}
    {\ensuremath \Cal{S}}

\newcommand{\SchwartzR}
    {\ensuremath \Schwartz (\R)}

\newcommand{\SchwartzRd}
    {\ensuremath \Schwartz (\R^d)}

\newcommand{\SchwartzDual}
    {\ensuremath \Cal{S}'}

\newcommand{\SchwartzRDual}
    {\ensuremath \Schwartz' (\R)}

\newcommand{\SchwartzRdDual}
    {\ensuremath \Schwartz' (\R^d)}

\newcommand{\HSNorm}[1]
    {\norm{#1}_{H^s(\R^2)}}

\newcommand{\HSNormA}[2]
    {\norm{#1}_{H^{#2}(\R^2)}}

\newcommand{\Holder}
    {H\"{o}lder }

\newcommand{\Holders}
    {H\"{o}lder's }

\newcommand{\Holderian}
    {H\"{o}lderian }

\newcommand{\HolderRNorm}[1]
    {\widetilde{\Vert}{#1}\Vert_r}

\newcommand{\LInfNorm}[1]
    {\norm{#1}_{L^\iny(\Omega)}}

\newcommand{\SmallLInfNorm}[1]
    {\smallnorm{#1}_{L^\iny}}

\newcommand{\LOneNorm}[1]
    {\norm{#1}_{L^1}}

\newcommand{\SmallLOneNorm}[1]
    {\smallnorm{#1}_{L^1}}

\newcommand{\LTwoNorm}[1]
    {\norm{#1}_{L^2(\Omega)}}

\newcommand{\SmallLTwoNorm}[1]
    {\smallnorm{#1}_{L^2}}

\newcommand{\LpNorm}[2]
    {\norm{#1}_{L^{#2}}}

\newcommand{\SmallLpNorm}[2]
    {\smallnorm{#1}_{L^{#2}}}

\newcommand{\lOneNorm}[1]
    {\norm{#1}_{l^1}}

\newcommand{\lTwoNorm}[1]
    {\norm{#1}_{l^2}}

\newcommand{\MsrNorm}[1]
    {\norm{#1}_{\Cal{M}}}

\newcommand{\FTF}
    {\Cal{F}}

\newcommand{\FTR}
    {\Cal{F}^{-1}}

\newcommand{\InvLaplacian}
    {\ensuremath{\widetilde{\Delta}^{-1}}}

\newcommand{\EqDef}
    {\hspace{0.2em}={\hspace{-1.2em}\raisebox{1.2ex}{\scriptsize def}}\hspace{0.2em}}

%
%

%
%

\title
    [On vanishing viscosity in a disk]
    {On the vanishing viscosity limit in a disk}

\author{James P. Kelliher}
\address{Department of Mathematics, Brown University, Box 1917, Providence, RI
         02912}
\curraddr{Department of Mathematics, Brown University, Box 1917, Providence, RI
          02912}
\email{kelliher@math.brown.edu}

\date{} 



\begin{abstract}
Let $u$ be a solution to the Navier-Stokes equations in the unit disk with
no-slip boundary conditions and viscosity $\nu > 0$, and let $\ol{u}$ be a
smooth solution to the Euler equations. We say that the vanishing viscosity
limit holds on $[0, T]$ if $u$ converges to $\ol{u}$ in $L^\iny([0, T]; L^2)$.
We show that a necessary and sufficient condition for the vanishing viscosity
limit to hold is the vanishing with the viscosity of the time-space average of
the energy of $u$ in a boundary layer of width proportional to $\nu$ due to the
modes (eigenfunctions of the Stokes operator) whose frequencies in the radial
or the tangential direction lie between $L(\nu)$ and $M(\nu)$. Here, $L(\nu)$
must be of order less than $1/\nu$ and $M(\nu)$ must be of order greater than
$1/\nu$.
\end{abstract}

\Obsolete{ 
    \begin{abstract}
        Let $\Omega$ be a bounded domain in $\R^d$, $d \ge 2$, and let
        $\Gamma_{\delta}$ be a layer of width $\delta$ around the boundary. Let $u$ be
        a solution to the Navier-Stokes equations on $\Omega$ with no-slip boundary
        conditions and viscosity $\nu > 0$, and let $\ol{u}$ be a smooth solution to
        the Euler equations on $\Omega$. We say that the vanishing viscosity limit
        holds on $[0, T]$ if $u$ converges to $\ol{u}$ in $L^\iny([0, T];
        L^2(\Omega))$. Various necessary and sufficient conditions on the solution $u$
        are known for the this limit to hold. Specializing to the case of a disk, we
        show that the vanishing viscosity limit holds if and only if the time-space
        average of the energy in a boundary layer of width proportional to $\nu$  of
        the solution of the Navier-Stokes equations due to modes (eigenfunctions of the
        Stokes operator) whose frequencies in the radial or the tangential direction
        lie between $L(\nu)$ and $M(\nu)$ vanishes with the vorticity. Here, $L(\nu)$
        is of order less than $1/\nu$ and $M(\nu)$ is of order greater than $1/\nu$.
    \end{abstract}
    }

\Obsolete{ 
    \begin{abstract}
        Let $\Omega$ be a bounded domain in $\R^d$, $d \ge 2$, and let
        $\Gamma_{\delta}$ be a layer of width $\delta$ around the boundary. Let $u$ be
        a solution to the Navier-Stokes equations on $\Omega$ with no-slip boundary
        conditions and viscosity $\nu > 0$, and let $\ol{u}$ be a smooth solution to
        the Euler equations on $\Omega$. We say that the vanishing viscosity limit
        holds on $[0, T]$ if $u$ converges to $\ol{u}$ in $L^\iny([0, T];
        L^2(\Omega))$. Various necessary and sufficient conditions on the solution $u$
        are known for the vanishing viscosity to hold. These conditions include the
        behavior of the $L^2$-norms of $u$, $\grad u$, or the vorticity ($\curl u$) on
        $\Gamma_{c \nu}$, or the tangential components of $\grad u$ on
        $\Gamma_{\delta}$ with $\delta/\nu \to \iny$ as $\nu \to 0$, as well as the
        behavior of the latter three quantities on all of $\Omega$. Specializing to the
        case of a disk, we show that in each of these conditions we need only consider
        contributions from modes (eigenfunctions of the Stokes operator) lying within
        certain ranges of frequencies, either high, low, or intermediate, depending
        upon the condition. In particular, we find that for conditions involving the
        vorticity in the whole domain or the velocity or its gradient in the boundary
        layer, one need only consider modes whose frequencies in the radial or the
        tangential direction lie between $L(\nu)$ and $M(\nu)$, where $L(\nu)$ is of
        order less than $1/\nu$ and $M(\nu)$ is of order greater than $1/\nu$.
    \end{abstract}
    }

\maketitle

\DetailMarginNote{
    \begin{small}
        \begin{flushright}
            Compiled on \textit{\textbf{\today}}

            \ExplainDetailLevel
        \end{flushright}
    \end{small}
    }


%
%
\section{Introduction}\label{S:Introduction}

\noindent In the presence of a boundary, the question of whether solutions of
the Navier-Stokes equations with no-slip boundary conditions converge to a
solution of the Euler equations as the viscosity vanishes---the so-called
vanishing viscosity limit---is very difficult. The convergence of most interest
is of the velocities, uniformly over finite time and $L^2$ in space. Except in
the very special case of radially symmetric initial vorticity in a disk, where
convergence is known to hold (see \refT{VVRadialH}), the question of
convergence or the lack thereof is unresolved for nonzero initial velocity in a
bounded domain. (For a half-space with analytic initial data, the vanishing
viscosity limit is shown to hold in \cite{SC1998II}.)

Tosio Kato in \cite{Kato1983} gave necessary and sufficient conditions on the
velocity $u$ of the Navier-Stokes equations for the vanishing viscosity limit
to hold. The most interesting of these is that
\begin{align*}
    \nu \int_0^T \norm{\grad u(t)}_{L^2(\Gamma_{c \nu})}^2 \, dt \to 0
        \text{ as } \nu \to 0,
\end{align*}
where $\Gamma_{c \nu}$ is the boundary strip of width $c \nu$ with $c > 0$
fixed but arbitrary. Making only a small change to Kato's proof, it is possible
to replace $\grad u$ with the vorticity $\omega = \omega(u) = \prt_1 u^2 -
\prt_2 u^1$, giving \refE{OmegaBoundaryLayerCondition} (see \cite{K2006Kato}).
(The necessity of \refE{OmegaBoundaryLayerCondition} is immediate from Kato's
condition, but because we do not have a boundary condition on the inner
boundary of $\Gamma_{c \nu}$ the sufficiency of the condition requires proof.)

Other necessary and sufficient conditions were established by Teman and Wang in
\cite{TW1998} and \cite{W2001}. These are the conditions in
\refE{TemamWangCondition1} and \refE{TemamWangCondition2} of
\refT{KatoKelliherTemamWang}, and involve only the derivatives in the
directions tangential to the boundary of either the tangential or normal
components of the velocity, though for a slightly larger boundary layer.
Finally, a condition that requires that the average energy density in the
boundary layer of the same width as Kato's vanish with viscosity,
\refE{uBoundaryLayerCondition}, is proven in \cite{K2006Kato}. All these
conditions (which apply to a bounded domain in dimensions 2 and higher) are
summarized in \refT{KatoKelliherTemamWang}.

We consider the issue of vanishing viscosity in the (unit) disk and look for
weaker necessary and sufficient conditions for the limit to hold. The reason
for working in the disk is that the simple geometry allows us to make quite
explicit calculations using the eigenfunctions of the Stokes operator, which
are composed of Bessel functions of the first kind. In a sense, this connects
the energy method with the geometry. What we find is that we need only consider
certain ranges of frequencies (or equivalently, length scales) in the various
conditions: this is \refT{MainResult}. Although \refT{MainResult} is specific
to the disk, there is no hydrodynamical reason to expect the disk to be special
as regards the vanishing viscosity limit, so one would expect a version of the
theorem to apply to all sufficiently smooth bounded domains in $\R^2$, and
probably in higher dimensions as well. We discuss this issue more fully in
\refR{MainResultInTermsOfEigenvalues}.

In \cite{CW2006}, Cheng and Wang obtain a result regarding vanishing viscosity
in two dimensions analogous to \refE{TemamWangCondition1New} and
\refE{TemamWangCondition2New}. Their result applies to an approximating
sequence to a solution of the Navier-Stokes equations as the viscosity
vanishes, whereas our result applies to the necessary and sufficient condition
for the vanishing viscosity limit to hold. While for the other conditions in
\refT{MainResult} we use very different techniques than those in \cite{CW2006},
our proof of the necessity and sufficiency of \refE{TemamWangCondition1New} and
\refE{TemamWangCondition2New} uses the key inequality in their paper.
\refS{CW2006} contains a brief comparison between the two results.

In \cite{CCS2001}, the authors consider the Stokes problem (linearized
Navier-Stokes equations) \textit{external} to a disk with time-varying
Dirichlet boundary conditions, showing that the vanishing viscosity limit
holds. In fact, they do much more than this, giving an explicit construction of
the solution to the Stokes problem and showing that it can be decomposed into
the sum of the solution to the linearized Euler equations, the solution to the
associated Prantdl equations, and a small correction term. The symmetry of the
geometry allows the authors of \cite{CCS2001} to construct the solutions in an
explicit form (involving Bessel functions of the first and second kind). The
nonlinear term in the Navier-Stokes equations makes an explicit solution
impossible for us; however, we can expand the solution in terms of
eigenfunctions of the Stokes operator for which we have an explicit form (in
terms of Bessel functions of the first kind) which we can use to obtain finer
estimates on the behavior of the Navier-Stokes equations in the boundary layer
than would be possible for a general domain.

A word on notation: We use $C$ to represent an unspecified constant that always
has the same value on both sides of an equality but may have a different value
on each side of an inequality.

\section{Definitions and {K}ato-type conditions}\label{S:Defs}

\noindent We now give definitions of the Euler and Navier-Stokes equations, and
state the results from \cite{Kato1983}, \cite{K2006Kato}, \cite{TW1998}, and
\cite{W2001} that we will need.

In \refS{EigenV} we will specialize to the unit disk, but for now we assume
only that $\Omega$ is a bounded domain in $\R^2$ with $C^2$-boundary $\Gamma$,
and we let $\mathbf{n}$ be the outward normal vector to $\Gamma$.

A classical solution $(\ol{u}, \ol{p})$ to the Euler equations satisfies, for
fixed $T > 0$,
\begin{align*}
    \begin{matrix}
        (E) & \left\{
            \begin{array}{l}
                \prt_t \ol{u} + \ol{u} \cdot \grad \ol{u} + \grad \ol{p} =
                \ol{f}
                        \text{ and }
                    \dv \ol{u} = 0 \text{ on } [0, T] \times \Omega, \\
                \ol{u}\cdot \mathbf{n} = 0 \text { on } [0, T] \times \Gamma,
                \text{ and } \ol{u} = \ol{u}^0 \text{ on } \set{0} \times
                \Omega,
            \end{array}
            \right.
    \end{matrix}
\end{align*}
where $\dv \ol{u}^0 = 0$. These equations describe the motion of an
incompressible fluid of constant density and zero viscosity.

We assume that $\ol{u}^0$ is in $C^{k + \eps}(\Omega)$, $\eps > 0$, and that
$\ol{f}$ is in $C^k([0, t] \times \Omega)$ for all $t
> 0$, where $k = 1$ or $2$. Then as shown in
\cite{Koch2002} (Theorem 1 and the remarks on p. 508-509), there exists a
unique solution $\ol{u}$ in $C^1_{loc}([0, \iny); C^{k + \eps}(\Omega))$.

The Navier-Stokes equations describe the motion of an incompressible fluid of
constant density and positive viscosity $\nu$. A classical solution to the
Navier-Stokes equations can be defined in analogy with ($E$) by
\begin{align*}
    \begin{matrix}
        & \left\{
            \begin{array}{l}
                \prt_t u + u \cdot \grad u + \grad p = \nu \Delta u + f
                        \text{ and }
                    \dv u = 0 \text{ on } [0, T] \times \Omega, \\
                u = 0 \text { on } [0, T] \times \Gamma,
                \text{ and } u = u_\nu^0 \text{ on } \set{0} \times \Omega.
            \end{array}
            \right.
    \end{matrix}
\end{align*}
We will work, however, with weak solutions to the Navier-Stokes equations.
\begin{definition}[Weak Navier-Stokes Solutions]\label{D:WeakSolutionNS}
    Given $T > 0$, viscosity $\nu > 0$, and initial velocity $u_\nu^0$ in $H$,
    $u$ in $L^2([0, T]; V)$ with $\prt_t u$ in $L^2([0, T]; V')$ is a weak solution to
    the Navier-Stokes equations if $u(0) = u_\nu^0$ and
    \begin{align*}
        (NS)
        \qquad \int_\Omega \prt_t u \cdot v
            + \int_\Omega (u \cdot \grad u) \cdot v
            + \nu \int_\Omega \grad u \cdot \grad v
            = \int_\Omega f v
    \end{align*}
    for all $v$ in $V$. (The spaces $H$ and $V$ are defined in
    \refS{StokesOperatorBD}.)
\end{definition}

\begin{definition}\label{D:VVLimit}
    We say that the \textit{vanishing viscosity limit holds} if
    \begin{align}\label{e:VVLimit}
        u \to \ol{u} \text{ in } L^\iny([0, T]; L^2(\Omega)) \text { as }
            \nu \rightarrow 0.
    \end{align}
\end{definition}

\refT{KatoKelliherTemamWang} applies to a bounded domain with $C^2$-boundary in
$\R^d$, $d \ge 2$. The conditions in \refE{OmegaL2Condition} and
\refE{GraduBoundaryLayerCondition} are due to Kato (\cite{Kato1983}), the
conditions in \refE{OmegaBoundaryLayerCondition} and
\refE{uBoundaryLayerCondition} appear in \cite{K2006Kato}, and the conditions
in \refE{TemamWangCondition1} and \refE{TemamWangCondition2} are due to Temam
and Wang (\cite{TW1998}, \cite{W2001}).

\begin{theorem}\label{T:KatoKelliherTemamWang}
    Let $T > 0$ and assume that $u_\nu^0$ is in $H$ and that $\ol{u}^0$ is in
    $C^{k + \eps}(\Omega)$, $\eps > 0$ with $k = 1$ or $2$. In addition, assume that
    \begin{itemize}
        \item[(a)]
            $u_\nu^0 \rightarrow \ol{u}^0$ in $L^2(\Omega)$
                as $\nu \rightarrow 0$,
        \item[(b)]
            $f$ is in $L^1([0, T]; L^2(\Omega))$,
        \item[(c)]
            $\smallnorm{f - \ol{f}}_{L^1([0, T]; L^2(\Omega))}
                \rightarrow 0$ as $\nu \rightarrow 0$.
    \end{itemize}
    Let $\delta: [0, \iny) \to [0, \iny)$ be such that $\delta(\nu)$
    converges to 0 while $\delta(\nu)/\nu$ diverges to $\iny$ as $\nu \to 0$.
    Then the the vanishing viscosity limit (\refD{VVLimit}) holds if and only
    if any of the following conditions holds:
    \begin{align}\label{e:OmegaL2Condition}
        \nu \int_0^T \norm{\omega(s)}_{L^2(\Omega)}^2 \, ds \to 0
            \text{ as } \nu \to 0,
    \end{align}
    \begin{align}\label{e:OmegaBoundaryLayerCondition}
        \nu \int_0^T \norm{\omega(s)}_{L^2(\Gamma_{c \nu})}^2 \, ds \to 0
            \text{ as } \nu \to 0,
    \end{align}
    \begin{align}\label{e:GraduBoundaryLayerCondition}
        \nu \int_0^T \norm{\grad u(s)}_{L^2(\Gamma_{c \nu})}^2 \, ds \to 0
            \text{ as } \nu \to 0,
    \end{align}
    \begin{align}\label{e:TemamWangCondition1}
        \nu \int_0^T \norm{\grad_{\BoldTau}
            u_{\BoldTau}(s)}_{L^2(\Gamma_{\delta(\nu)})}^2 \, ds \to 0
                \text{ as } \nu \to 0,
    \end{align}
    \begin{align}\label{e:TemamWangCondition2}
        \nu \int_0^T \norm{\grad_{\BoldTau}
            u_{\mathbf{n}}(s)}_{L^2(\Gamma_{\delta(\nu)})}^2 \, ds \to 0
                \text{ as } \nu \to 0.
    \end{align}
    Here $\grad_{\BoldTau}$ represents the derivatives in the boundary layer in the
    directions tangential to the boundary, $u_{\BoldTau}$ is the projection of $u$ in the
    direction tangential to the boundary, and $u_{\mathbf{n}}$ is the projection of $u$ in the
    direction normal to the boundary.

    When $k = 2$, these conditions are also equivalent to
    \begin{align}\label{e:uBoundaryLayerCondition}
        \frac{1}{\nu} \int_0^T \norm{u(s)}_{L^2(\Gamma_{c \nu})}^2 \, ds \to 0
            \text{ as } \nu \to 0.
    \end{align}
\end{theorem}

The quantity in \refE{uBoundaryLayerCondition} is proportional to the
space-time average of the energy in the boundary layer.

We show (see \refR{GenHighFrequencyResult}) that in \refE{OmegaL2Condition},
\refE{GraduBoundaryLayerCondition}, and \refE{uBoundaryLayerCondition},
contributions from the high frequency modes can be ignored. This result applies
to an arbitrary bounded domain in $\R^d$, $d \ge 2$, with a $C^2$-boundary.

Our main result is \refT{MainResult}, which is an improvement of
\refT{KatoKelliherTemamWang} in the special case of the unit disk. In what
follows we decompose the solution $u$ in the form
\begin{align*}
    u(t, x) = \sum_{m = 0}^\iny \sum_{j = 1}^\iny
            g_{mj}(t) u_{mj}(x),
\end{align*}
where $(u_{mj})$ are the eigenfunctions of the Stokes operator described in
\refS{StokesOperatorBD} and \refS{EigenV}, and let
\begin{align}\label{e:uN}
    u^N(t, x) = \sum_{m = 0}^N \sum_{j = 1}^N
            g_{mj}(t) u_{mj}(x)
\end{align}
and
\begin{align}\label{e:uTildeN}
    \widetilde{u}^N(t, x) = \sum_{m = 0}^N \sum_{j = 1}^\iny
            g_{mj}(t) u_{mj}(x)
\end{align}
with vorticities $\omega^N(t, x) = \omega(u^N(t, x))$ and
$\widetilde{\omega}^N(t, x) = \omega(\widetilde{u}^N(t, x))$.

As we will see in \refS{EigenV}, the frequency of $u_{mk}$ in the tangential
direction is $m$ and the radial frequency of $u_{mk}$ is, in effect, $k$. Thus,
$u^N$ includes the contributions from all modes with both frequencies less than
$N$, while $\widetilde{u}^N$ includes the contributions from all modes with
tangential frequency less than $N$.

\begin{theorem}\label{T:MainResult}
    Assume that $\Omega$ is the unit disk and make the same assumptions on the
    initial data, forcing, and the function $\delta$ as in
    \refT{KatoKelliherTemamWang}. Let $L$ and $M$ be any functions mapping
    $(0, \iny)$ to $\Z^+$ with
    \begin{align}\label{e:LMNCondition}
        \nu L(\nu) \to 0, \;
        \nu M(\nu) \to \iny \text{ as } \nu \to 0.
    \end{align}

    Then the the vanishing viscosity limit (\refD{VVLimit}) holds if and only
    if any of the following conditions holds:
    \begin{align}\label{e:OmegaL2ConditionNew1}
        \nu \int_0^T \smallnorm{\omega(s)^{M(\nu)} - \omega^{L(\nu)}(s)}_{L^2(\Omega)}^2
            \, ds \to 0 \text{ as } \nu \to 0,
    \end{align}
    \begin{align}\label{e:OmegaL2ConditionNew2}
        \nu \int_0^T \smallnorm{\omega(s) - \widetilde{\omega}^{L(\nu)}(s)}_{L^2(\Omega)}^2
            \, ds \to 0 \text{ as } \nu \to 0,
    \end{align}
    \begin{align}\label{e:OmegaBoundaryLayerConditionNew}
        \nu \int_0^T \smallnorm{\omega(s)  - \omega^{L(\nu)}(s)}_{L^2(\Gamma_{c \nu})}^2
                \, ds \to 0 \text{ as } \nu \to 0,
    \end{align}
    \begin{align}\label{e:GraduBoundaryLayerConditionNew}
        \nu \int_0^T \smallnorm{\grad u^{M(\nu)}(s)
                - \grad u^{L(\nu)}(s)}_{L^2(\Gamma_{c \nu})}^2 \, ds \to 0
                    \text{ as } \nu \to 0,
    \end{align}
    \begin{align}\label{e:TemamWangCondition1New}
        \nu \int_0^T \smallnorm{\grad_{\BoldTau} u_{\BoldTau}(s)
                - \grad_{\BoldTau} \widetilde{u}^{L(\delta)}_{\BoldTau}(s)}_{L^2(\Gamma_{\delta(\nu)})}^2
                     \, ds \to 0 \text{ as } \nu \to 0,
    \end{align}
    \begin{align}\label{e:TemamWangCondition2New}
        \nu \int_0^T \smallnorm{\grad_{\BoldTau} u_{\mathbf{n}}(s)
                - \grad_{\BoldTau} \widetilde{u}^{L(\delta)}_{\mathbf{n}}(s)}_{L^2(\Gamma_{\delta(\nu)})}^2
                \, ds \to 0 \text{ as } \nu \to 0.
    \end{align}

    When $k = 2$, these conditions are also equivalent to
    \begin{align}\label{e:uBoundaryLayerConditionNew}
        \frac{1}{\nu} \int_0^T \smallnorm{u^{M(\nu)}(s) - u^{L(\nu)}(s)}_{L^2(\Gamma_{c \nu})}^2
            \, ds \to 0 \text{ as } \nu \to 0.
    \end{align}
\end{theorem}

Observe, for instance, that $u^{M(\nu)} - u^{L(\nu)}$ in
\refE{uBoundaryLayerConditionNew} represents the contribution from all modes
whose frequencies in the radial or the tangential direction lie between
$L(\nu)$ and $M(\nu)$.

\begin{remark}\label{R:MainResultInTermsOfEigenvalues}
    By \refL{jnkRange} and \refE{Eigenvalues}, $u^N$ is essentially the
    contributions of all the modes with eigenvalues less than $C N^2$. In fact,
    suppose that we replace the definition of $u^N$ in \refE{uN} with
    \begin{align}\label{e:uNAlt}
        u^N(t, x) = \sum_{\set{j: \la_j < N^2}}
                g_{j}(t) u_{j}(x),
    \end{align}
the single subscripts in \refE{uNAlt} referring to the eigenfunctions and
eigenvalues of the Stokes operator on a general domain in $\R^d$, $d \ge 2$,
defined in \refS{StokesOperatorBD}. It follows easily from \refT{MainResult}
that the conditions in \refE{OmegaL2ConditionNew1},
\refE{OmegaBoundaryLayerConditionNew}, \refE{GraduBoundaryLayerConditionNew},
and \refE{uBoundaryLayerConditionNew} continue to be equivalent to the
vanishing viscosity limit. It is in this form that we would expect
\refT{MainResult} to generalize to fairly arbitrary smooth domains in $\R^2$
and---with $N^2$ in \refE{uNAlt} replaced by $N$ raised to some other
power---to domains in $\R^d$, $d \ge 3$. The obstacle to establishing this
generalization is the difficulty of obtaining the equivalents of
\refL{L2omegaGammaBoundGeneral} and \refL{L2uGammaBoundGeneral}---along with an
approximate form of \refL{SomeL2InnerProductsAreZero}---for high frequencies.
\end{remark}

\section{The Stokes operator in a bounded domain}\label{S:StokesOperatorBD}

\noindent Before specializing to the case of a disk, we discuss first some
general properties related to the Stokes operator.

We define the function spaces $H$ and $V$ as follows (see Section I.1.4 of
\cite{T2001} for more details). First let
\begin{align*}
    \Cal{V} = \set{u \in (\Cal{D}(\Omega))^2: \dv u = 0}
\end{align*}
be the space of vector-valued divergence-free distributions on $\Omega$. We let
$H$ be the closure of $\Cal{V}$ in $L^2(\Omega)$ and $V$ be the closure of
$\Cal{V}$ in $H^1_0(\Omega)$. Alternate characterizations of $H$ and $V$ are
\begin{align*}
    H &= \set{u \in (L^2(\Omega))^2: \dv u = 0 \text{ in } \Omega, \,
                u \cdot \mathbf{n} = 0 \text{ on } \Gamma}, \\
    V &= \set{u \in (L^2(\Omega))^2: \dv u = 0 \text{ in } \Omega, \,
                u = 0 \text{ on } \Gamma},
\end{align*}
the boundary conditions applying in terms of a trace.

By $\innp{\cdot, \cdot}$ we mean the inner product in $L^2(\Omega)$: $\innp{f,
g} = \int_\Omega f \ol{g}$. (It will be convenient to use complex-valued
eigenfunctions, so the complex conjugate is required in this definition. Our
velocity fields and vorticities, however, are real, so conjugation will not
always appear in our calculations.) Then $\innp{u, v}_H = \innp{u, v}$ and
$\innp{u, v}_V = \innp{\grad u, \grad v}$.

Although $\Cal{V}$ is dense in $H$ it is not dense in $H \cap H^1(\Omega)$
(with the $H^1$-norm). Informally, this is because each element of $\Cal{V}$ is
zero on $\Gamma$ and so the limit of a sequence of elements in $\Cal{V}$ cannot
become nonzero on the boundary without the gradient near the boundary becoming
indefinitely large. More formally, we have \refL{NotDenseInH}.

\begin{lemma}\label{L:NotDenseInH}
    The space $\Cal{V}$ is not dense in $H \cap H^1(\Omega)$.
\end{lemma}
\begin{proof}
    Let $u$ be any element of $V$. Then its vorticity $\omega$ is in $L^2(\Omega)
    \subseteq L^1(\Omega)$ and must satisfy
    \begin{align}\label{e:OmegaZero}
        \int_\Omega \omega
        &= \int_\Omega \Delta \psi
            = \int_\Omega \dv \grad \psi
            = \int_\Gamma \grad \psi \cdot \mathbf{n}
            = - \int_\Gamma u^\perp \cdot \mathbf{n}
            = 0,
    \end{align}
    where $\psi$ is the stream function: $u = \grad^\perp \psi = (-\prt_2 \psi,
    \prt_1 \psi)$ and $\omega = \Delta \psi$. Because only $u \cdot \mathbf{n} = 0$
    for $u$ in $H$, the same cannot be said for $u$ in $H \cap H^1(\Omega)$: Let
    $u$ in $H \cap H^1(\Omega)$ have vorticity $\omega$ with nonzero total mass.
    Then for any sequence $\set{v_j}$ in $\Cal{V}$,
    \begin{align*}
        \norm{\omega - \omega(v_j)}_{L^2(\Omega)}
        &\ge C \norm{\omega - \omega(v_j)}_{L^1(\Omega)}
            \ge C \abs{\int_\Omega (\omega - \omega(v_j))} \\
        &= C \abs{\int_\Omega \omega}
            > 0,
    \end{align*}
    so $\Cal{V}$ cannot be dense in $H \cap H^1(\Omega)$.
\end{proof}

We now briefly describe the properties we will need of the Stokes operator $A$
on $\Omega$, referring the reader, for instance, to Section I.2 of \cite{T2001}
for more details. One way to define $A$ is that given $u$ in $V \cap
H^2(\Omega)$, $Au$ in $H$ satisfies $Au = - \Delta u + \grad p$ for some
harmonic scalar field $p$. We have $D(A) = V \cap H^2(\Omega)$ with $A$ mapping
$D(A)$ onto $H$, and there exists a set of eigenfunctions $\set{u_j}$ for $A$,
complete in $H$ and in $V$, with corresponding eigenvalues $\set{\lambda_j}$,
$0 < \lambda_1 \le \lambda_2 \le \cdots$, and each $u_j$ is in $H^2(\Omega)$
since we are assuming that $\Gamma$ is $C^2$. (When we specialize to the disk,
the eigenfunctions will be in $C^\iny(\Omega)$.) An eigenfunction $u_j$ of $A$
satisfies $A u_j = \la_j u_j$ or, equivalently,
\begin{align}\label{e:StokesEigenfunctionSolution}
    \begin{matrix}
        \left\{
            \begin{array}{l}
                \Delta u_j + \la_j u_j = \grad p_j, \, \Delta p_j = 0, \, \dv u_j = 0,
                    \text{ on } \Omega, \\
                    u_j = 0 \text{ on } \Gamma.
            \end{array}
            \right.
    \end{matrix}
\end{align}
The eigenfunctions are orthogonal in both $H$ and $V$. The usual convention is
to make the eigenvectors orthonormal in $H$, but we will find it more
convenient to normalize them to be orthonormal in $V$ so that $\norm{\grad
u_j}_{L^2(\Omega)}^2 = \norm{\omega_j}_{L^2(\Omega)}^2 = 1$ and
\begin{align}\label{e:ujNorm}
    \norm{u_j}_{L^2(\Omega)}^2
       &= \innp{u_j, u_j}
        = \frac{1}{\la_j} \innp{u_j, A u_j}
        = \frac{1}{\la_j} \innp{\grad u_j, \grad u_j}
        = \frac{1}{\la_j}.
\end{align}
Moreover, we have \refL{DenseInV}.

\begin{lemma}\label{L:DenseInV}
    If $u$ is in $V$ with $\omega = \omega(u)$ then
    \begin{align}\label{e:HExpansion}
        u = \sum_{j = 1}^\iny \innp{\omega, \omega_j} u_j,
    \end{align}
    with the sum converging in both $V$ and $H$.
\end{lemma}
\begin{proof}
    Let $u$ be in $V$ and let $u^n = \sum_{j = 1}^n (\innp{u, u_j}_H/\innp{u_j, u_j}_H) u_j$. Then
    $u^n$ converges in $H$ to $u$ because $\set{u_j}$ is complete in $H$.
    But,
    \begin{align*}
        \frac{\innp{u, u_j}_H}{\innp{u_j, u_j}_H}
           &= \frac{\la_j \innp{u, u_j}}{\la_j \innp{u_j, u_j}}
            = \frac{\innp{u, A u_j}}{\innp{u_j, A u_j}}
            = \frac{\innp{\grad u, \grad u_j}}{\innp{\grad u_j, \grad u_j}} \\
           &= \innp{\grad u, \grad u_j}
            = \innp{\omega, \omega_j},
    \end{align*}
    so the expansion of $u$ in $V$ in terms of the eigenfunctions of $A$ is the
    same as the expansion of $u$ in $H$ (and the coefficients are as given in
    \refE{HExpansion}), meaning that $u^n$ converges in $V$ to $u$ as well.
\end{proof}

In the proof of \refL{DenseInV} we used the identity $\innp{\grad u, \grad v} =
\innp{\omega(u), \omega(v)}$ for all $u$, $v$ in $V$, which follows by
integrating by parts. Were we to use the definition of $\omega(u)$ as the
antisymmetric matrix $(\grad u - (\grad u)^T)/2$, which is usual in higher
dimensions, this would have introduced a factor of 2 into \refE{HExpansion}.

\begin{cor}\label{C:uomegaExpansion}
    If $u$ is in $V$ then
    \begin{align*}
        \grad u = \sum_{j = 1}^\iny \innp{\omega, \omega_j} \grad u_j
                \text{ and }
            \omega = \sum_{j = 1}^\iny \innp{\omega, \omega_j} \omega_j,
    \end{align*}
    with the sums converging in $L^2(\Omega)$.
\end{cor}

\Obsolete{
    It\MarginNote{Numerical evidence seems to suggest that it might,
    however, although one more derivative does not.}is important to bear in mind
    that for $u$ in $V \cap H^2(\Omega)$, the expansion in \refE{HExpansion} need
    not converge in $V \cap H^2(\Omega)$, even though $\Cal{V} \cap H^2(\Omega)$ is
    dense in $V \cap H^2(\Omega)$. This is because the eigenfunctions $\set{u_j}$
    are not orthogonal in $V \cap H^2(\Omega)$.
    }

Since the solution $u$ to ($NS$) lies in $V$ for all positive time, we can
write
\begin{align}\label{e:SingleOmegaExpansion}
    \begin{split}
        \omega(t) = \sum_{j = 1}^\iny g_j(t) \omega_j&, \quad
            u(t) = \sum_{j = 1}^\iny g_j(t) u_j, \\
        \norm{\omega(t)}_{L^2(\Omega)}^2
                = \sum_{j = 1}^\iny \abs{g_j(t)}^2&, \quad
            \norm{u(t)}_{L^2(\Omega)}^2
                = \sum_{j = 1}^\iny \frac{\abs{g_j(t)}^2}{\la_j},
    \end{split}
\end{align}
where $g_j$ are functions of time. The expansion of $u$ will converge for all
$t \ge 0$ and that of $\omega$ for $t > 0$---and also for $t = 0$ if and only
if the initial velocity is in $V$; in general, we only assume that it is $H$.
Because $u(t) \to u_\nu^0$ in $L^2(\Omega)$ as $t \to 0$, each $g_j(t)$ is
continuous at $t = 0$, though this does not mean that $\omega(t)$ is continuous
in $L^2(\Omega)$ at $t = 0$. Also, note that $g_j(t)$ is complex-valued since
the eigenvectors are complex-valued, but $u(t)$ and $\omega(t)$ are
real-valued.

\section{Eigenfunctions of the Stokes operator in the unit disk}\label{S:EigenV}

\noindent We now fix $\Omega$ to be the unit disk in $\R^2$ centered at the
origin.

In \cite{LR2002}, a complete set of eigenfunctions for the annulus is derived
in terms of Bessel functions of the first and second kind, $J_n$ and $Y_n$. By
ignoring the terms involving $Y_n$ and modifying somewhat the calculation of
the eigenvalues, one can easily obtain the eigenfunctions for a disk. We will,
however, derive the vorticity of the eigenfunctions directly, as this is quite
easy. In determining the eigenvalues and the velocity of the eigenfunctions,
which is more difficult, we will rely on the results in \cite{LR2002}.

Taking the curl of \refE{StokesEigenfunctionSolution} (with $u = u_j$), we see
that the vorticity $\omega = \omega(u)$ satisfies
\begin{align}\label{e:OmegaStokesEigenfunctionSolution}
    \begin{matrix}
        \left\{
            \begin{array}{l}
                \Delta \omega + \la \omega = 0
                    \text{ on } \Omega, \\
                    u = 0 \text{ on } \Gamma.
            \end{array}
            \right.
    \end{matrix}
\end{align}
That is, $\omega$ is an eigenfunction  of the negative Laplacian, but with
boundary conditions on the velocity $u$.

Ignoring for the moment the issue of boundary conditions, we use separation of
variables to look for a complete set of solutions to $\Delta \omega + \la
\omega = 0$ on $\Omega$. Writing
\begin{align*}
    \omega(r, \theta) = f_n(r) e^{i n \theta}
\end{align*}
in polar coordinates, $n = 0, 1, 2, \dots$, and using
\begin{align*}
    \Delta = \pdxtwo{}{r} + \frac{1}{r} \pdx{}{r}
        + \frac{1}{r^2} \pdxtwo{}{\theta}
\end{align*}
gives
\begin{align}\label{e:fnBasicEquation}
    \pr{\difftwo{f_n}{r} + \frac{1}{r} \diff{f_n}{r}
            + \pr{\la - \frac{n^2}{r^2}} f_n}
        e^{i n \theta} = 0.
\end{align}

Since $J_n$, the Bessel function of the first kind of order $n$, is a solution
of
\begin{align}\label{e:BesselDE}
    \difftwo{J_n(s)}{s} + \frac{1}{s} \diff{J_n(s)}{s}
            - \pr{1 - \frac{ n^2}{s^2}} J_n(s)
        = 0,
\end{align}
making the change of variables $s = \la^{1/2} r$, we see that
\refE{fnBasicEquation} holds with $f_n(r) = J_n(\lambda^{1/2} r)$. Thus, the
eigenfunctions have vorticity of the form $J_n(\lambda^{1/2}  r) e^{i n
\theta}$ and it remains to determine the eigenvalues that satisfy $u = 0$ on
the boundary.

The easiest way to do this is to use the expressions in \cite{LR2002}. For $n =
0$, we drop the term involving $Y_1$ in Equation (30) p. 406 of \cite{LR2002},
giving
\begin{align*}
    u_{0k}(r, \theta) = \lambda_{0k}^{-1/2}
        J_1(\lambda_{0k}^{1/2} r) \wh{e}_\theta,
\end{align*}
where $\lambda_{0k}^{1/2}$, $k = 1, 2, \dots$, are the eigenvalues described
below. For $n \ge 1$, dropping the terms involving the Bessel functions of the
second kind from the last equation on p. (406) of \cite{LR2002}, we have
\begin{align*}
    u_{nk}(r, \theta)
       &= \pr{\frac{i n}{\lambda_{nk} r} J_n(\la_{nk}^{1/2} r)
            + \frac{D_{nk} i n^2}{\lambda_{nk}^2} r^{n - 1}}
                e^{i n \theta} \wh{e}_r \\
           &
            + \pr{\frac{1}{2 \la_{nk}^{1/2}}
                \pr{J_{n + 1}(\la_{nk}^{1/2} r)
                        - J_{n - 1}(\la_{nk}^{1/2} r)}
                - \frac{D_{nk} n^2}{\lambda_{nk}^2} r^{n - 1}}
                e^{i n \theta} \wh{e}_\theta,
\end{align*}
the eigenvalues $\lambda_{nk}$, $k = 1, 2, \dots$, being described below and
the $D_{nk}$ being undetermined constants. In both cases we scaled the
eigenfunctions differently than in \cite{LR2002}. A direct calculation shows
that
\begin{align*}
    \omega_{nk}(r, \theta)
        \EqDef \omega(u_{nk})(r, \theta)
        = C_{nk} J_n(\la_{nk}^{1/2} r) e^{i n \theta},
\end{align*}
where $C_{nk}$ is a normalization constant. A direct calculation also shows
that $\dv u_{nk} = 0$.

For $n = 0$, $\lambda_{0k}^{1/2} = j_{1k}$, where
\begin{align}\label{e:jnk}
    j_{nk} \text{ is the }k\textit{-th} \text{ positive root of }
        J_{n + 1}(x) = 0,
\end{align}
as this gives $u_{0k}(1, \theta) = 0$. Setting $u_{nk}(1, \theta) = 0$ we
obtain two equations in the two unknowns $\lambda_{nk}$ and $D_{nk}$. We
eliminate $D_{nk}$ from the two equations to obtain a single equation for
$\lambda_{nk}$. Then using the identity in \refE{Bowman629} we obtain the
equation
\begin{align*}
    \la_{nk}^{1/2} J_n'(\la_{nk}^{1/2})
        - n J_n(\la_{nk}^{1/2}) = 0.
\end{align*}
Thus, $\la_{nk}^{1/2}$ is the $k$-th positive root of
\begin{align}\label{e:EigenRoot}
    x J_n'(x) - n J_n(x) = -x J_{n + 1}(x) = 0,
\end{align}
$k = 1, 2, \dots$, where we used \refE{Bowman631}. That is, $\la_{nk}^{1/2} =
j_{n + 1, k}$. It follows then that
\begin{align}\label{e:Dnk}
    D_{nk}
        = - \frac{\la_{nk} J_n(\la_{nk}^{1/2})}{n}
        = -\frac{j_{n + 1, k}^2 J_n(j_{n + 1, k})}{n}.
\end{align}

Since we are normalizing the eigenfunctions so that $\innp{\omega_{mj},
\omega_{nk}} = \delta_{mn} \delta_{jk}$, we must choose $C_{nk}$ so that
\begin{align*} 
    \begin{split}
        C_{nk}^{-2}
        &= \smallnorm{J_n(j_{n + 1, k} r) e^{i n \theta}}_{L^2(\Omega)}^2
         = 2 \pi \int_0^1 r J_n(j_{n + 1, k} r)^2 \, dr \\
        &= 2 \pi \frac{r^2}{2} \brac{J_n(j_{n + 1, k} r)^2
                    - J_{n - 1}(j_{n + 1, k} r) J_{n + 1}(j_{n + 1, k} r)}_0^1 \\
        &= \pi J_n(j_{n + 1, k})^2.
    \end{split}
\end{align*}
Here we used \refE{Bowman653}.

To summarize, the vorticity of the eigenfunctions is given by
\begin{align*}
    \omega_{nk}(r, \theta)
        = C_{nk} J_n(j_{n + 1, k} r) e^{i n \theta},
\end{align*}
with eigenvalue
\begin{align}\label{e:Eigenvalues}
    \la_{nk} = j_{n +1, k}^2,
\end{align}
and where
\begin{align*}
    C_{nk} = \frac{1}{\pi^{1/2} \smallabs{J_n(j_{n + 1, k})}},
\end{align*}
$n = 0, 1, \dots$, $k = 1, 2, \dots$. With our choice of normalization of the
eigenfunctions (\refE{ujNorm}), the velocity becomes
\begin{align}\label{e:unk}
    \begin{split}
        u_{nk}&(r, \theta)
            = \frac{J_n(\al r) - J_n(\al) r^n}
                {\pi^{1/2} \al^2 \abs{J_n(\al)} r}
                        i n e^{i n \theta} \wh{e}_r \\
            &
                + \frac{\al \pr{J_{n + 1}(\al r)
                            - J_{n - 1}(\al r)}
                            + 2 n J_n(\al) r^{n - 1}}
                    {2 \pi^{1/2} \al^2 \abs{J_n(\al)}}
                    e^{i n \theta} \wh{e}_\theta,
    \end{split}
\end{align}
where $\al = j_{n + 1, k}$.

\Obsolete{
    The expressions including $\omega$ in \refE{SingleOmegaExpansion}
    become
    \begin{align}\label{e:OmegaExpansion}
        \begin{split}
            \omega(t, r, \theta)
            &= \sum_{n = 0}^\iny \sum_{k = 1}^\iny g_{nk}(t) \omega_{nk}(r)
                = \sum_{n = 0}^\iny e^{i n \theta}
                    \sum_{j = 1}^\iny g_{nj}(t) C_{nj} J_n(j_{n + 1, j} r), \\
            \norm{\omega(t)}_{L^2(\Omega)}
            &= \sum_{n = 0}^\iny
                    \sum_{j = 1}^\iny g_{nj}(t)^2
        \end{split}
    \end{align}
    for all $t > 0$. The third sum is, of course, absolutely convergent and so
    independent of the ordering of the eigenfunctions; the first two sums are
    independent of the ordering of the eigenfunctions almost everywhere in
    $\Omega$. Hence, it is not important to describe how we are to view these
    double sums as a limit of partial sums.
    }

\Obsolete{
    We can describe the eigenfunctions $\omega_{nk}$ semi-quantitatively
    as follows.

    The absolute value of each eigenfunction is constant on the boundary, the value
    in each case being $\pi^{-1/2}$. The maximum value of $\abs{\omega_{nk}}$
    occurs at the first positive zero of $J_n'(j_{n + 1, k} r)$, which lies between
    the first and second positive zeros of $J_n(j_{n + 1, k} r)$; that is, between
    $j_{n, 1}/j_{n + 1, k}$ and $j_{n, 2}/j_{n + 1, k}$. Because $\abs{J_n(x)} \le
    C x^{-1/2}$ and because of \refL{jnkRange} and \refL{Jnalphabound}, the
    $L^\iny$-norm of $\omega_{nk}$ is of order $((n + k)/n)^{1/2} = (1 +
    k/n)^{1/2}$.

    As $n$ gets large, $J_n$ is quite near zero until shortly before its first
    positive zero, where it reaches it maximum. Thus, $\omega_{n, 1}$ has its mass
    concentrated farther toward the boundary as $n$ becomes large. For fixed $n$,
    as $k$ increases, the mass of $\abs{\omega_{nk}}$ starts to move in toward the
    origin, goes from near zero to its maximum value very quickly, then decays
    roughly like $C r^{-1/2}$, oscillating through $k$ zeros on its way to its
    value of $\pi^{-1/2}$ on the boundary. As $k$ becomes very large relative to
    $n$, the mass of $\omega_{nk}$ moves in to approach the origin, its
    $L^\iny$-norm approaches infinity (like $(1 + k/n)^{1/2}$, as observed above),
    and there becomes a larger and larger difference between the magnitude of
    $\omega_{nk}$ near the origin and on the boundary.

    And, of course, the phase of $\omega_{nk}$ goes through $n$ periods around the
    disk.
    }

\Obsolete{
    Since the eigenvalues are distinct, if $m \ne n$ or $j \ne k$,
    \begin{align*}
    &\innp{\grad u_{mj}, \grad u_{nk}}_{L^2(\Omega)}
            = - \innp{u_{mj}, \Delta u_{nk}}_{L^2(\Omega)} \\
        &\qquad\qquad = \innp{u_{mj}, A u_{nk} + \grad p}_{L^2(\Omega)}
            = \lambda_{nk} \innp{u_{mj}, u_{nk}}_{L^2(\Omega)}
    \end{align*}
    and similarly $\innp{\grad u_{mj}, \grad u_{nk}}_{L^2(\Omega)} = \lambda_{mj}
    \innp{u_{mj}, u_{nk}}_{L^2(\Omega)}$. We conclude that $\innp{\grad u_{mj},
    \grad u_{nk}}_{L^2(\Omega)} = 0$. Since
    \begin{align*}
        \innp{\grad u, \grad v}_{L^2(\Omega)}
            = \innp{\omega(u), \omega(v)}_{L^2(\Omega)}
    \end{align*}
    for all $u$, $v$ in $V$, we conclude that $\innp{\omega_{mj},
    \omega_{nk}}_{L^2(\Omega)} = 0$. This also follows directly from properties of
    the Bessel functions; we recognize, for instance, the inner sum in
    \refE{OmegaExpansion} as a time-varying Dini expansion (see, for instance, p.
    108-110 of \cite{Bowman}).

    \textbf{TODO}: Add a short comment about the asymptotics of the eigenvalues and
    how they agree with what we would expect.
    }

\Obsolete{
    \section{Eigenfunctions for the Stokes operator in $H$}\label{S:EigenH}

    \noindent \textbf{Eventually make this section obsolete, although it contains
    some ideas that might eventually be useful. The terminology is also not the
    best, calling something an eigenfunction for the Stokes operator in $H$.}

    \skipline

    \noindent It is easier to determine the eigenfunctions for $A$ in $H$, because
    we can take advantage of the stream function. For any $u$ in $H \cap C^\iny$
    there exists a \textit{stream function} $\psi$ such that $u = \grad^\perp \psi
    = (-\prt_2 \psi, \prt_1 \psi)$. Furthermore, $\Delta \psi = \omega$. The
    boundary conditions for $u$ in $H$ are satisfied by $\psi = 0$ on $\Gamma$, so
    given $\omega$, $\psi$ satisfies
    \begin{align}\label{e:OmegaPsi}
        \begin{matrix}
            \left\{
                \begin{array}{ll}
                    \Delta \psi = \omega & \text{on } \Omega, \\
                        \psi = 0 & \text{on } \Gamma.
                \end{array}
                \right.
        \end{matrix}
    \end{align}

    As shown in \refS{EigenV},  $\omega(r, \theta) = J_n(\lambda^{1/2} r) e^{i n
    \theta}$ for some nonnegative integer $n$ and some eigenvalue $\la$, and
    $\Delta \omega = - \la \omega$. Thus letting
    \begin{align*}
        \psi
            = -(1/\la) \omega
            =  -(1/\la) J_n(\lambda^{1/2} r) e^{i n \theta}
    \end{align*}
    will satisfy \refE{OmegaPsi} as long as we choose $\lambda^{1/2}$ to be a
    positive zero of $J_n$. Because $\Omega$ is simply connected, $\psi$
    identically zero gives $u$ identically zero, so $\lambda = 0$ must be rejected
    for these solutions. The eigenvalue of zero, however, corresponds to the
    solution $\omega = 1$ in \refE{OmegaStokesEigenfunctionSolution}, which in turn
    corresponds to $\psi(r, \theta) = (1/4) (r^2 - 1)$ in \refE{OmegaPsi}, which
    has a unique solution. Thus, we must also include a constant eigenfunction.

    (The reason we exclude the eigenvalue zero in \refS{EigenV} is because for any
    $u$ in $V$ the vorticity $\omega$ must satisfy
    \begin{align*}
        \int_\Omega \omega
        &= \int_\Omega \Delta \psi
            = \int_\Omega \dv \grad \psi
            = \int_\Gamma \grad \psi \cdot \mathbf{n}
            = - \int_\Gamma u^\perp \cdot \mathbf{n}
            = 0,
    \end{align*}
    so $\omega = 1$ has no associated velocity lying in $V$. Or because $(Au, u) =
    (\grad u, \grad u) > 0$ is $u \ne 0$; that is, $A$ is positive-definite.)

    Thus, a complete set of eigenfunctions for the vorticity
    $\set{\ol{\omega}_{nk}}$ with associated eigenvalues $\set{\mu_{nk}}$ is given
    by
    \begin{align*}
        \left\{
            \begin{array}{l}
                \ol{\omega}_{nk}(r, \theta)
                    = J_n(\mu_{nk}^{1/2} r) e^{i n \theta},
                        \; n = 0, 1, \dots, \, k = 1, 2, \dots, \\
                \mu^{1/2}
                    = k\textit{-th} \text{ positive root of }
                        J_n(x) = 0,
            \end{array}
        \right.
    \end{align*}
    along with the constant function $1$ (or we could include $\lambda_{00} = 1$).
    Thus, we can write the initial vorticity $\omega^0$ as
    \begin{align*} 
        \omega^0(r, \theta)
        &= h_{00}+ \sum_{n = 0}^\iny e^{i n \theta}
                \sum_{k = 1}^\iny h_{nk} J_n(\mu_{nk}^{1/2} r) .
    \end{align*}
    }

\section{Proof of \refT{MainResult}}\label{S:ProofOfMainResult}

\noindent From the fundamental energy equality for ($NS$) we have for all $t$
in $[0, T$],
\begin{align*}
    \nu \int_0^t \norm{\grad u}_{L^2(\Omega)}^2
        = \nu \int_0^t \norm{\omega}_{L^2(\Omega)}^2
        \le \frac{1}{2} \smallnorm{u_\nu^0}_H^2
            + 4 \norm{f}_{L^1([0, T]; L^2(\Omega))}^2.
\end{align*}
It follows from \refE{SingleOmegaExpansion} and assumptions (a) and (b) of
\refT{KatoKelliherTemamWang} that for all sufficiently small $\nu > 0$,
\begin{align}\label{e:ajConvergence}
    \begin{split}
        \nu \int_0^t \norm{\omega}_{L^2(\Omega)}^2
           &= \nu \int_0^t \sum_{m = 0}^\iny \sum_{j = 1}^\iny \abs{g_{mj}(s)}^2 \, ds
            \le C.
    \end{split}
\end{align}

\begin{theorem}\label{T:VanishingVorticity}
    With the assumptions of \refT{MainResult},
    \begin{align}\label{e:LowFrequencyConvergence}
        \lim_{\nu \to 0}
            \nu \int_0^t \smallnorm{\omega^{L(\nu)}}_{L^2(\Gamma_{c \nu})}^2
            = 0.
    \end{align}
\end{theorem}
\begin{proof}
    Using \refL{SomeL2InnerProductsAreZero},
    \begin{align*}
        \nu \int_0^t &\smallnorm{\omega^{L(\nu)}}_{L^2(\Gamma_{c \nu})}^2 \\
            &= \nu \int_0^t \sum_{m = 0}^{L(\nu)} \sum_{j = 1}^{L(\nu)}
                            \sum_{n = 0}^{L(\nu)} \sum_{k = 1}^{L(\nu)}
                g_{mj}(s) \ol{g_{nk}(s)} \, ds
                    \innp{\omega_{mj}, \omega_{nk}}_{L^2(\Gamma_{c \nu})}
    \end{align*}
    \vspace{-6pt}
    \begin{align*}
        \hspace{-1pt}
                &= \nu \int_0^t \sum_{n = 0}^{L(\nu)} \sum_{j = 1}^{L(\nu)} \sum_{k = 1}^{L(\nu)}
                g_{nj}(s) \ol{g_{nk}(s)} \, ds
                    \innp{\omega_{nj}, \omega_{nk}}_{L^2(\Gamma_{c \nu})}
    \end{align*}
    \vspace{-6pt}
    \begin{align*}
        \qquad\qquad\quad
            &\le \nu \int_0^t \sum_{n = 0}^{L(\nu)} \sum_{j = 1}^{L(\nu)} \sum_{k = 1}^{L(\nu)}
                \abs{g_{nj}(s)} \abs{g_{nk}(s)} \, ds
                    \norm{\omega_{nj}}_{L^2(\Gamma_{c \nu})}
                    \norm{\omega_{nk}}_{L^2(\Gamma_{c \nu})}
    \end{align*}
    \vspace{-6pt}
    \begin{align*}
        \hspace{-19pt}
            &= \nu \int_0^t \sum_{n = 0}^{L(\nu)}
                    \pr{\sum_{j = 1}^{L(\nu)} \abs{g_{nj}(s)}
                    \norm{\omega_{nj}}_{L^2(\Gamma_{c \nu})}}^2 \, ds
    \end{align*}
    \vspace{-6pt}
    \begin{align*}
        \hspace{-17pt}
            &\le \nu \int_0^t \sum_{n = 0}^{L(\nu)}
                    \sum_{j = 1}^{L(\nu)} \abs{g_{nj}(s)}^2
                    \sum_{j = 1}^{L(\nu)}  \norm{\omega_{nj}}_{L^2(\Gamma_{c \nu})}^2 \,
                        ds,
    \end{align*}
    where we used the Cauchy-Schwarz inequality in the last step.

    \Obsolete{ 
        \begin{align*}
            \nu \int_0^t &\smallnorm{\omega^{L(\nu)}}_{L^2(\Gamma_{c \nu})}^2 \\
                &= \nu \int_0^t \sum_{m = 0}^{L(\nu)} \sum_{j = 1}^{L(\nu)}
                                \sum_{n = 0}^{L(\nu)} \sum_{k = 1}^{L(\nu)}
                    g_{mj}(s) \ol{g_{nk}(s)} \, ds
                        \innp{\omega_{mj}, \omega_{nk}}_{L^2(\Gamma_{c \nu})} \\
                &= \nu \int_0^t \sum_{n = 0}^{L(\nu)} \sum_{j = 1}^{L(\nu)} \sum_{k = 1}^{L(\nu)}
                    g_{nj}(s) \ol{g_{nk}(s)} \, ds
                        \innp{\omega_{nj}, \omega_{nk}}_{L^2(\Gamma_{c \nu})} \\
                &\le \nu \int_0^t \sum_{n = 0}^{L(\nu)} \sum_{j = 1}^{L(\nu)} \sum_{k = 1}^{L(\nu)}
                    \abs{g_{nj}(s)} \abs{g_{nk}(s)} \, ds
                        \norm{\omega_{nj}}_{L^2(\Gamma_{c \nu})}
                        \norm{\omega_{nk}}_{L^2(\Gamma_{c \nu})} \\
                &= \nu \int_0^t \sum_{n = 0}^{L(\nu)}
                        \pr{\sum_{j = 1}^{L(\nu)} \abs{g_{nj}(s)}
                        \norm{\omega_{nj}}_{L^2(\Gamma_{c \nu})}}^2 \, ds \\
                &\le \nu \int_0^t \sum_{n = 0}^{L(\nu)}
                        \sum_{j = 1}^{L(\nu)} \abs{g_{nj}(s)}^2
                        \sum_{j = 1}^{L(\nu)}  \norm{\omega_{nj}}_{L^2(\Gamma_{c \nu})}^2 \,
                            ds,
        \end{align*}
        }

    By \refL{jnkRange},
    \begin{align*}
        1/L(\nu)
            < C/(L(\nu) + 2)
            \le C/j_{L(\nu) + 1, 1}
            = C\la_{L(\nu), 1}^{-1/2}.
    \end{align*}
    Since $\nu L(\nu) \to 0$ as $\nu \to 0$, for all sufficiently small $\nu$ we
    have $c \nu < C/L(\nu) \le 2 \pi \la_{L(\nu), 1}^{-1/2} \le 2 \pi \la_{j, 1}^{-1/2}$
    for all $j \le L(\nu)$. So by \refL{L2omegaGammaBoundGeneral},
    \begin{align}\label{e:InnerSumBoundN}
        \sum_{j = 1}^{L(\nu)}  \norm{\omega_{nj}}_{L^2(\Gamma_{c \nu})}^2
        &\le C \nu L(\nu).
    \end{align}

    Then using \refE{ajConvergence},
    \begin{align*}
        \nu \int_0^t &\smallnorm{\omega^{L(\nu)}}_{L^2(\Gamma_{c \nu})}^2
            \le C \nu L(\nu)  \pr{\nu \int_0^t \sum_{n = 0}^{L(\nu)}
                    \sum_{j = 1}^{L(\nu)} \abs{g_{nj}(s)}^2
                        ds}
            \le C \nu L(\nu),
    \end{align*}
    which vanishes with $\nu$ by the condition in \refE{LMNCondition}, and
    \refE{LowFrequencyConvergence} therefore holds.
\end{proof}

\begin{remark}\label{R:TryImproveOmega}
    We could try to improve \refT{VanishingVorticity} by using
    $\omega(\widetilde{u}^N)$ of \refE{uTildeN} in place of $\omega^N$, thereby
    incorporating all of the frequencies in the radial direction for a given
    angular frequency. Unfortunately, the best bound that one can achieve on
    $\norm{\omega_{nj}}_{L^2(\Gamma_{\delta})}^2$ for $j > n$ is the extension of
    \refL{L2omegaGammaBoundGeneral} described in \refR{L2omegaGammaBoundGeneral},
    and this is very much insufficient to bound the terms with $j > n$.

    Another possible approach is to try to incorporate the destructive interference
    that occurs in the inner product of two eigenfunctions in the boundary layer
    that the use of \Holders inequality in our proof of \refT{VanishingVorticity}
    ignored. The best bound one can hope to obtain is that
    \begin{align*}
        \smallabs{\innp{\omega_{nj}, \omega_{nk}}_{L^2(\Gamma_{\delta})}}
            \le \frac{C}{\abs{k - j}}
    \end{align*}
    for all $\delta$ in $[0, 1]$ and without restriction on $n$, $j$, or $k$ except
    that $j \ne k$. We could then follow the obvious approach of decomposing the
    equivalent of the first sum in the proof of \refT{VanishingVorticity} into four
    pieces: a diagonal term where $m = n$ and $j = k$ and three terms containing
    low frequencies in $j$ and $k$, low frequencies in $j$ and high frequencies in
    $k$, and high frequencies in both $j$ and $k$. If we do this, however, we will
    find that the factor of $1/\abs{k - j}$ is just insufficient to obtain
    convergence.
\end{remark}

\Obsolete{ 
    The fundamental obstacle is that we do not know, and should not expect, that
    the sum that defines $\omega(\widetilde{u}^N)$ converges absolutely. Thus,
    unless there is some technique that treats the inner sum over $j$ as an
    entirety, it is hard to see how we can improve \refT{VanishingVorticity}. This
    is in contrast to the conditions in \refE{TemamWangCondition1New} and
    \refE{TemamWangCondition2New}, which as we will see in the proof of
    \refT{VanishingutangentialResult} follow very easily from a type of reverse
    Poincar\'{e} inequality, as employed in \cite{CW2006}.
    }

\begin{cor}\label{C:MainCorOmega}
    The conditions in \refE{VVLimit} and \refE{OmegaBoundaryLayerConditionNew} of
    \refT{MainResult} are equivalent.
\end{cor}
\begin{proof}
    That \refE{VVLimit} implies \refE{OmegaBoundaryLayerConditionNew} follows
    directly from \refT{KatoKelliherTemamWang}. So assume that
    \refE{OmegaBoundaryLayerConditionNew} holds. Because $\norm{A + B}^2 \le 2
    \norm{A}^2 + 2 \norm{B}^2$ for any norm,
        \begin{align*}
            &\nu \int_0^t \smallnorm{\omega}_{L^2(\Gamma_{c \nu})}^2
                \le 2 \nu \int_0^t \smallnorm{\omega^{L(\nu)}}_{L^2(\Gamma_{c \nu})}^2
                + 2 \nu \int_0^t \smallnorm{\omega - \omega^{L(\nu)}}_{L^2(\Gamma_{c
                            \nu})}^2.
        \end{align*}
    This vanishes with $\nu$ by \refT{VanishingVorticity} and
    \refE{OmegaBoundaryLayerConditionNew}, showing that
    \refE{OmegaBoundaryLayerCondition} holds and hence by
    \refT{KatoKelliherTemamWang} that \refE{VVLimit} holds.
\end{proof}

\begin{theorem}\label{T:VanishinguResult}
    With the assumptions of \refT{MainResult},
    \begin{align}\label{e:HighFrequencyuConvergence}
        \lim_{\nu \to 0}
            \frac{1}{\nu} &\int_0^t \smallnorm{u(s) - u^{M(\nu)}(s)}_{L^2(\Gamma_{c
                \nu})}^2 \, ds
            = 0
    \end{align}
    and
    \begin{align}\label{e:LowFrequencyuConvergence}
        \lim_{\nu \to 0}
            \frac{1}{\nu} \int_0^t
                \smallnorm{u^{L(\nu)}(s)}_{L^2(\Gamma_{c \nu})}^2 \, ds
            = 0.
    \end{align}
\end{theorem}
\begin{proof}
    We can write $u(t) - u^{M(\nu)}(t) = A(t) + B(t)$, where
    \begin{align*}
        A(t) = \sum_{m = 1}^{M(\nu)} \sum_{j = M(\nu) + 1}^\iny g_{mj}(t) u_{mj}(x),
            \quad
        B(t) = \hspace{-12pt}
            \sum_{m = M(\nu) + 1}^\iny \sum_{j = 1}^\iny g_{mj}(t) u_{mj}(x)
    \end{align*}
    and
    \begin{align*}
        \smallnorm{u(t) - u^{M(\nu)}(t)}_{L^2(\Gamma_{c \nu})}^2
           &\le 2\norm{A(t)}_{L^2(\Gamma_{c \nu})}^2
                + 2\norm{B(t)}_{L^2(\Gamma_{c \nu})}^2.
    \end{align*}
    Now,
    \begin{align*}
       &\smallnorm{A(t)}_{L^2(\Gamma_{c \nu})}^2
            \le \smallnorm{A(t)}_{L^2(\Omega)}^2
            = \sum_{m = 1}^{M(\nu)} \sum_{j = M(\nu) + 1}^\iny
                \abs{g_{mj}(t)}^2 \norm{u_{mj}}_{L^2(\Omega)}^2 \\
           &\qquad = \sum_{m = 1}^{M(\nu)} \sum_{j = M(\nu) + 1}^\iny
                \frac{\abs{g_{mj}(t)}^2}{\la_{mj}}
            \le \frac{1}{\la_{1M(\nu)}} \sum_{m = 1}^{M(\nu)} \sum_{j = M(\nu) + 1}^\iny
                \abs{g_{mj}(t)}^2 \\
           &\qquad \le \frac{1}{\la_{1M(\nu)}} \smallnorm{\omega(t) -
                   \omega^{M(\nu)}(t)}_{L^2(\Omega)}^2,
    \end{align*}
    where we used \refE{ujNorm}. Similarly,
    \begin{align*}
       &\smallnorm{B(t)}_{L^2(\Gamma_{c \nu})}^2
            \le \smallnorm{B(t)}_{L^2(\Omega)}^2
            \le \frac{1}{\la_{M(\nu)1}} \smallnorm{\omega(t) -
                   \omega^{M(\nu)}(t)}_{L^2(\Omega)}^2.
    \end{align*}
    By \refE{Eigenvalues} and \refL{jnkRange}, $\la_{M(\nu)1}$ and $\la_{1M(\nu)}$
    are both bounded below (and above) by $CM(\nu)^2$, so
    \begin{align*}
        \smallnorm{u(t) - u^{M(\nu)}(t)}_{L^2(\Gamma_{c \nu})}^2
           &\le \frac{C}{M(\nu)^2} \smallnorm{\omega(t)
                - \omega^{M(\nu)}(t)}_{L^2(\Omega)}^2.
    \end{align*}
    \Obsolete{
        (This is really just Poincar\'{e} inequality with the constant being the
        lowest eigenvalue in the eigenfunction decomposition of $u - u^{M(\nu)}$ in $\Omega$.)
        }

    Then,
    \begin{align*}
        \frac{1}{\nu} &\int_0^t \smallnorm{u(s) - u^{M(\nu)}(s)}_{L^2(\Gamma_{c
                \nu})}^2 \, ds \\
           &\le \frac{C}{\nu M(\nu)^2} \int_0^t
                \smallnorm{\omega(s) - \omega^{M(\nu)}(s)}_{L^2(\Omega)}^2 \,
                        ds \\
           &\le \frac{C}{\nu M(\nu)^2} \int_0^t
                \smallnorm{\omega(s)}_{L^2(\Omega)}^2 \,
                        ds \\
           &= \frac{C}{\nu^2 M(\nu)^2} \nu \int_0^t
                    \smallnorm{\omega(s)}_{L^2(\Omega)}^2 \,
                        ds
            \le \frac{C}{\nu^2 M(\nu)^2},
    \end{align*}
    where in the last inequality we used \refE{ajConvergence}. This vanishes
    with $\nu$ by the assumption on $M$ in \refE{LMNCondition} giving
    \refE{HighFrequencyuConvergence}.

    Arguing as in the proof of
    \refT{VanishingVorticity},
    \begin{align*}
        \frac{1}{\nu} &\int_0^t
                \smallnorm{u^L(s)}_{L^2(\Gamma_{c \nu})}^2 \, ds
            \le \frac{1}{\nu}
                \int_0^t \sum_{n = 0}^{L(\nu)}
                    \sum_{j = 1}^{L(\nu)} \abs{g_{nj}(s)}^2
                    \sum_{j = 1}^{L(\nu)}  \norm{u_{nj}}_{L^2(\Gamma_{c \nu})}^2 \,
                        ds \\
           &\le \frac{C L(\nu) \nu^3}{\nu}
                \int_0^t \sum_{n = 0}^{L(\nu)}
                    \sum_{j = 1}^{L(\nu)} \abs{g_{nj}(s)}^2
                        ds
            \le C L(\nu) \nu
    \end{align*}
    for all sufficiently small $\nu$.
    In the second inequality we used \refL{L2uGammaBoundGeneral} and in the last
    inequality we used \refE{ajConvergence}. This integral also vanishes with $\nu$
    by the assumption on $L$ in \refE{LMNCondition} giving
    \refE{LowFrequencyuConvergence}.
\end{proof}

\begin{cor}\label{C:MainCoru}
    The conditions in \refE{VVLimit}, \refE{OmegaL2ConditionNew1},
    \refE{GraduBoundaryLayerConditionNew}, and \refE{uBoundaryLayerConditionNew} of
    \refT{MainResult} are equivalent.
\end{cor}
\begin{proof}
    For sufficiently large $\nu$, $L(\nu) \le M(\nu)$, and we have
    \begin{align*}
       &\smallnorm{u(s)}_{L^2(\Gamma_{c \nu})}^2
            \le 3 \smallnorm{u^{M(\nu)}(s) - u^{L(\nu)}(s)}_{L^2(\Gamma_{c \nu})}^2
                 \\
           &\qquad
                + 3 \smallnorm{u^{L(\nu)}(s)}_{L^2(\Gamma_{c \nu})}^2
                + 3 \smallnorm{u(s) - u^{M(\nu)}(s)}_{L^2(\Gamma_{c \nu})}^2.
    \end{align*}
    It follows from \refT{VanishinguResult} that
    \begin{align*}
        \limsup_{\nu \to 0} \frac{1}{\nu}
            \int_0^t \smallnorm{u(s)}_{L^2(\Gamma_{c \nu})}^2
           &\le
        3 \limsup_{\nu \to 0} \frac{1}{\nu}
            \int_0^t \smallnorm{u^{M(\nu)} - u^{L(\nu)}}_{L^2(\Gamma_{c
                \nu})}^2.
    \end{align*}
    In particular, the first limsup is zero if and only if the second limsup is
    zero (the reverse inequality without the factor of 3 being trivial). Then
    \refE{uBoundaryLayerCondition} of \refT{KatoKelliherTemamWang} shows that
    \refE{uBoundaryLayerConditionNew} holds if and only if \refE{VVLimit} holds.
    The sufficiency of \refE{OmegaL2ConditionNew1} and
    \refE{GraduBoundaryLayerConditionNew} for \refE{VVLimit} to hold then follows
    from Poincar\'{e}'s inequality in the form
    \begin{align*}
       &\smallnorm{u^{M(\nu)}(s) - u^{L(\nu)}(s)}_{L^2(\Gamma_{c \nu})}^2
            \le C \nu^2 \smallnorm{\grad u^{M(\nu)}(s) - \grad u^{L(\nu)}(s)}_{L^2(\Gamma_{c \nu})}^2
                \\
          &\qquad \le C \nu^2 \smallnorm{\grad u^{M(\nu)}(s) - \grad u^{L(\nu)}(s)}_{L^2(\Omega)}^2
                \\
          &\qquad= C \nu^2 \smallnorm{\omega^{M(\nu)}(s) - \omega^{L(\nu)}(s)}_{L^2(\Omega)}^2.
    \end{align*}
    The necessity of \refE{OmegaL2ConditionNew1} and
    \refE{GraduBoundaryLayerConditionNew} follow immediately from
    \refT{KatoKelliherTemamWang}.
\end{proof}

\begin{remark}\label{R:GenHighFrequencyResult}
If we replace the definition of $u^N$ in \refE{uN} with that in \refE{uNAlt},
then it is clear that \refE{HighFrequencyuConvergence} continues to hold in any
bounded domain in $\R^2$ with a $C^2$-boundary. It follows as in
\refC{MainCoru} that the vanishing viscosity limit of \refD{VVLimit} holds if
and only if the condition in \refE{OmegaL2ConditionNew1},
\refE{GraduBoundaryLayerConditionNew}, or (when $k = 2$)
\refE{uBoundaryLayerConditionNew} holds with the term involving $u^{L(\nu)}$ in
each of these conditions removed. A similar result would hold in any dimension
for an arbitrary bounded domain with a $C^2$-boundary.
\end{remark}

\begin{theorem}\label{T:VanishingutangentialResult}
    With the assumptions of \refT{MainResult},
    \begin{align}\label{e:HighFrequencyTemamWangConvergence1}
        \nu \int_0^T \smallnorm{\grad_{\BoldTau}
                \widetilde{u}^{L(\delta)}_{\BoldTau}(s)}_{L^2(\Gamma_{\delta(\nu)})}^2
                     \, ds \to 0 \text{ as } \nu \to 0
    \end{align}
    and
    \begin{align}\label{e:HighFrequencyTemamWangConvergence2}
        \nu \int_0^T \smallnorm{\grad_{\BoldTau}
            \widetilde{u}^{L(\delta)}_{\mathbf{n}}(s)}_{L^2(\Gamma_{\delta(\nu)})}^2
                \, ds \to 0 \text{ as } \nu \to 0.
    \end{align}
\end{theorem}
\begin{proof}
    In the unit disk, $u_{\BoldTau} = u^\theta$ and $\grad_{\BoldTau} =
    \prt_\sigma$, where $\sigma$ is arc length along the circle of radius $r$, in
    which $r$ is held constant. Thus,
    \begin{align*}
        \grad_{\BoldTau} u_{\BoldTau}
            = \pdx{u^\theta}{\sigma}
            = \frac{1}{r} \pdx{u^\theta}{\theta}
    \end{align*}
    and for any positive integer $N$
    it follows from Poincar\'{e}'s inequality that
    \begin{align*}
       &\smallnorm{\grad_{\BoldTau}
                \widetilde{u}^N_{\BoldTau}(s)}_{L^2(\Gamma_\delta)}^2
            = \norm{\frac{1}{r} \pdx{}{\theta}
                (\widetilde{u}^N(s))^\theta}_{L^2(\Gamma_\delta)}^2
            \le \frac{1}{(1 - \delta)^2} \norm{\pdx{}{\theta}
                (\widetilde{u}^N(s))^\theta}_{L^2(\Gamma_\delta)}^2 \\
            &\qquad
                \le \frac{C \delta^2}{(1 - \delta)^2} \norm{\frac{\prt^2}{\prt r \prt \theta}
                    (\widetilde{u}^N(s))^\theta}_{L^2(\Gamma_\delta)}^2
                \le \frac{C \delta^2}{(1 - \delta)^2} \norm{\frac{\prt^2}{\prt r \prt \theta}
                   (\widetilde{u}^N(s))^\theta}_{L^2(\Omega)}^2.
    \end{align*}

    But,
    \begin{align*}
        \frac{\prt^2}{\prt r \prt \theta}
                (\widetilde{u}^N(s))^\theta
        &= \sum_{m = 0}^N \sum_{j = 1}^\iny
                g_{mj}(s) \frac{\prt^2}{\prt r \prt \theta} u_{mj}(r, \theta) \\
        &= i \sum_{m = 0}^N m \sum_{j = 1}^\iny
                g_{mj}(s) \pdx{}{r} u_{mj} (r, \theta),
    \end{align*}
    the last equality following from the simple dependence of $u_{mj}$ on $\theta$
    in \refE{unk}. Thus,
    \begin{align*}
       &\norm{\frac{\prt^2}{\prt r \prt \theta}
                (\widetilde{u}^N(s))^\theta}_{L^2(\Omega)}^2
            \le \norm{i \sum_{m = 0}^N m \sum_{j = 1}^\iny
                g_{mj}(s) \grad u_{mj} (r, \theta)}_{L^2(\Omega)}^2 \\
          &\qquad
            = \sum_{m = 0}^N m^2 \sum_{j = 1}^\iny \abs{g_{mj}(s)}^2
            \le N^2 \sum_{m = 0}^N \sum_{j = 1}^\iny \abs{g_{mj}(s)}^2
            \le N^2 \smallnorm{\grad u}_{L^2(\Omega)}^2,
    \end{align*}
    where we used the orthonormality of the eigenfunctions in $V$.

    Combining these two inequalities gives
    \begin{align*}
        \smallnorm{\grad_{\BoldTau}
                \widetilde{u}^N_{\BoldTau}(s)}_{L^2(\Gamma_\delta)}^2
            \le \frac{C N^2 \delta^2}{(1 - \delta)^2} \smallnorm{\grad u}_{L^2(\Omega)}^2.
    \end{align*}

    Then using \refE{ajConvergence},
    \begin{align*}
        \nu \int_0^T \smallnorm{\grad_{\BoldTau}
                \widetilde{u}^{L(\delta)}_{\BoldTau}(s)}_{L^2(\Gamma_{\delta(\nu)})}^2
        &\le \frac{C L(\delta)^2 \delta^2}{(1 - \delta)^2}
                \nu \int_0^T \smallnorm{\grad u}_{L^2(\Omega)}^2
            \le \frac{C L(\delta)^2 \delta^2}{(1 - \delta)^2}.
    \end{align*}
    This vanishes with $\delta$ by the assumption on $L$ in \refE{LMNCondition}
    and hence vanishes with $\nu$ since $\delta$ vanishes with $\nu$, giving
    \refE{HighFrequencyTemamWangConvergence1}. The proof of
    \refE{HighFrequencyTemamWangConvergence2} is entirely analogous.
\end{proof}

The technique used in the proof of \refT{VanishingutangentialResult} comes from
the key inequality following Equation (3.21) in \cite{CW2006}.

\begin{cor}\label{C:MainCorTemamWang}
    The conditions in \refE{VVLimit}, \refE{OmegaL2ConditionNew2},
    \refE{TemamWangCondition1New}, and
    \refE{TemamWangCondition2New} of \refT{MainResult} are equivalent.
\end{cor}
\begin{proof}
    This corollary can be proved much along the lines of the proofs of
    \refC{MainCorOmega} and \refC{MainCoru}. (It is here that we use the
    assumption that $\delta(\nu)/\nu$ diverges to $\iny$ as $\nu \to 0$,
    which is needed in applying \refT{KatoKelliherTemamWang}.)
\end{proof}

Together, \refC{MainCorOmega}, \refC{MainCoru}, and \refC{MainCorTemamWang}
establish \refT{MainResult}.

\section{Radially symmetric initial vorticity}\label{S:RadV}

\noindent It follows  from \refL{NotDenseInH} that if an initial velocity, no
matter how smooth, lies in $H$ but not in $V$ and has a vorticity whose total
mass is nonzero, then the velocity of the corresponding solution to ($NS$) will
be discontinuous in $H^1(\Omega)$ at time zero. In the same way, the vanishing
viscosity limit of the vorticity cannot hold in $L^\iny([0, T]; L^2(\Omega))$
for such an initial velocity, since the total mass of the vorticity for the
solution to ($E$) is conserved over time. This means that we might expect a
different character to the vanishing viscosity limit when the initial vorticity
has zero total mass versus when it has nonzero total mass.

Indeed, this is what happens in the special case of radially symmetric initial
vorticity where, when the initial velocity is in $V$ (which is equivalent for
radially symmetric vorticity to the total mass of the vorticity being zero) we
obtain convergence in $L^\iny([0, T]; L^2(\Omega))$ of both the velocity and
the vorticity (see \cite{BW2002}), whereas for initial velocity in $H$ we
obtain only the convergence of the velocity in this space, as in
\refT{VVRadialH}. Such convergence follows immediately from the conditions in
\refE{OmegaL2ConditionNew2}, \refE{TemamWangCondition1New}, or
\refE{TemamWangCondition2New} of \refT{MainResult}. The convergence also
follows from the sufficiency of the conditions in \refE{TemamWangCondition1}
and \refE{TemamWangCondition2} as established in \cite{TW1998}, since both
conditions are satisfied (the gradients in the tangential direction being zero)
as pointed out in \cite{W2001}. When the forcing is zero, however, there is a
simple proof that uses only Kato's original conditions.

\begin{theorem}\label{T:VVRadialH}
    Assume that $u_\nu^0$ and $\ol{u}^0$ are as in \refT{KatoKelliherTemamWang}
    with (for simplicity) $u_\nu^0 = \ol{u}^0$,
    that $f = \ol{f} = 0$, and that $\omega^0 = \omega(u^0)$ is
    radially symmetric. Then the vanishing viscosity limit of
    \refE{VVLimit} holds.
\end{theorem}
\begin{proof}
    Because $\omega^0$ is radially symmetric, $\omega$ remains radially symmetric
    for all time, so $\omega(u \cdot \grad u) = u \cdot \grad \omega = 0$. Then
    because $\Omega$ is simply connected, $u \cdot \grad u = \grad q$ for some
    scalar field $q$, and the nonlinear term in ($NS$) disappears. Thus, ($NS$)
    reduces to $u_\nu(0) = \ol{u}^0$ and
    \begin{align}\label{e:Heat}
        \int_\Omega \prt_t u_\nu \cdot v
            + \nu \int_\Omega \grad u_\nu \cdot \grad v
            = 0
    \end{align}
    for all $v$ in $V$. This is the heat equation in weak form, which is invariant
    under the transformation $(\nu, t, x) \mapsto (1, \nu t, x)$. That is, if $u_1$
    is a solution to \refE{Heat} with $\nu = 1$, then $u_\nu(t, x) = u_1(\nu t, x)$
    is a solution to \refE{Heat} because $u_\nu(0) = u_1(0) = \ol{u}^0$ and
    \begin{align*}
        \int_\Omega &\pdx{}{t} u_1(\nu t, x) \cdot v(x) \, dx
                + \nu \int_\Omega \grad u_1(\nu t, x) \cdot \grad v(x) \, dx \\
        &= \nu \brac{\int_\Omega (\prt_t u_1)(\nu t, x) \cdot v(x) \, dx
                + \int_\Omega \grad u_1(\nu t, x) \cdot \grad v(x) \, dx}
            = 0.
    \end{align*}

    It follows that
    \begin{align*}
        \nu \int_0^t \norm{\omega(s)}_{L^2(\Omega)}^2 \, ds
        &= \nu \int_0^t \norm{\omega_1(\nu s)}_{L^2(\Omega)}^2 \, ds
            = \int_0^{\nu t} \norm{\omega_1(\tau)}_{L^2(\Omega)}^2 \, d \tau.
    \end{align*}
    This vanishes as $\nu \to 0$ by the continuity of the integral, because $u_1$
    is in $L^2([0, T]; V)$. The limit in \refE{VVLimit} then follows from
    the condition in \refE{OmegaL2Condition} of \refT{KatoKelliherTemamWang}.
\end{proof}

The proof of of \refT{VVRadialH} does not yield a bound on the rate of
convergence in \refE{VVLimit}. Also, without assuming that the initial
vorticity is radially symmetric, the argument in the proof of \refT{VVRadialH}
can be applied to solutions to the Stokes problem (the linearized Navier-Stokes
equations) to show that they converge in the vanishing viscosity limit to a
solution to the linearized Euler equations (which is just the steady state
solution $\ol{u} = \ol{u}^0$). This would be more interesting, though, if
time-varying Dirichlet boundary conditions, for instance, could be
incorporated, as in \cite{CCS2001}.

In the simpler case of $u^0_\nu$ also lying in $V$, the solution to ($E$),
which is steady state, is zero on the boundary. This eliminates the troublesome
boundary term that appears in the direct energy argument bounding $\norm{u(t) -
\ol{u}(t)}_{L^2(\Omega)}$, giving an extremely simple proof of \refE{VVLimit}.
We give, however, a longer proof of convergence using
\refT{KatoKelliherTemamWang}, because it suggests how we might treat more
general initial velocities. For convenience, we assume zero forcing and more
regularity on the initial velocity than is strictly necessary.

\begin{theorem}\label{T:VVRadialV}
    Assume that $\omega_\nu^0$ is radially symmetric, $u^0_\nu = \ol{u}^0$
    is in $H^3(\Omega) \cap V$, and there is no forcing. Then the vanishing
    viscosity limit of \refE{VVLimit} holds.
\end{theorem}
\begin{proof}
    Because $\ol{u}^0$ is in $V \cap H^3(\Omega)$, $u$ is in $L^\iny_{loc}([0, \iny); V
    \cap H^3(\Omega))$ (see, for instance, Theorem III.3.6, Remark III.3.7, and
    Theorem III.3.10 of \cite{T2001}).

    As observed in the proof of \refT{VVRadialH}, $u \cdot \grad u = \grad q$ for
    some scalar field $q$. Then $\prt_t u + \grad (p + q) = \nu \Delta u$ and
    taking the divergence of both sides we conclude that $\Delta (p + q) = 0$.
    Because of the radial symmetry, however, $p + q$ is constant on $\Gamma$ and
    hence is constant on $\Omega$. Thus, $\grad (p + q) = 0$ and $\prt_t u
    = \nu \Delta u$. But $u = 0$ on $\Gamma$ so $\prt_t u = 0$ on $\Gamma$, and it
    follows that $\Delta u = \nu^{-1} \prt_t u$ is in $V$.

    Because $u$ is in $V$ at time zero, we can use the expansion for $\omega$ in
    \refE{SingleOmegaExpansion} for all time, including for time zero. Because $\omega$
    remains radially symmetric over time, the expansion reduces to
    \begin{align}\label{e:OmegaRadialExpansion}
        \omega(t, r, \theta)
            &= \sum_{k = 1}^\iny g_{0k}(t) \omega_{0k}(r)
            = \sum_{k = 1}^\iny g_{0k}(t) C_{0k} J_0(j_{1k} r).
    \end{align}
    Since $\Delta u$ is in $V$, it follows from \refC{uomegaExpansion} that $\Delta
    \omega$ has an expansion like that of \refE{OmegaRadialExpansion}:
    \begin{align}\label{e:DeltaOmegaRadialExpansion}
        \Delta \omega(t, r, \theta)
            = \sum_{k = 1}^\iny h_{0k}(t) C_{0k} J_0(j_{1k} r).
    \end{align}
    (The analogous expansion of $\Delta \omega$ including all the eigenvectors
    $\set{\omega_{nk}}$ fails to converge at $t = 0$ for non-radially symmetric
    solutions because $\Delta u$ is not, in general, in $V$.)

    Since $\Delta J_0(j_{1k} r) = - \lambda_{0k} J_0(j_{1k} r)$, it follows that
    $h_{0k}(t) = - \la_{0k}(t) g_{0k}(t)$ and then by \refE{Heat} in strong
    vorticity form,
    \begin{align}\label{e:VorticityFormulationE}
        \prt_t \omega = \nu \Delta \omega,
    \end{align}
    that $g_{0k}'(t) = - \nu \la_{0k} g_{0k}(t)$. Thus, $g_{0k}(t) = g_{0k}(0)
    e^{-\nu \lambda_{0k} t}$ and
    \begin{align}\label{e:HeatSolution}
        \omega(t, r, \theta)
            &= \sum_{k = 1}^\iny C_{0k} g_{0k}(0) e^{-\nu \lambda_{0k} t}
                J_0(j_{1k} r).
    \end{align}

    But then
    \begin{align}\label{e:L2RadVortBound}
        \begin{split}
            &\norm{\omega(t)}_{L^2(\Omega)}^2
                    = \sum_{k = 1}^\iny C_{0k}^2 g_{0k}(0)^2 e^{-2 \nu \lambda_{0k} t}
                        \smallnorm{J_0(j_{1k} r)}_{L^2(\Omega)}^2 \\
                &\qquad \le \sum_{k = 1}^\iny g_{0k}(0)^2
                        \smallnorm{C_{0k} J_0(j_{1k} r)}_{L^2(\Omega)}^2
                    = \smallnorm{\omega^0}_{L^2(\Omega)}^2
        \end{split}
    \end{align}
    and convergence in the vanishing viscosity limit follows from the condition in
    \refE{OmegaL2Condition} of \refT{KatoKelliherTemamWang}. (Examining the
    argument in \cite{Kato1983} shows that the convergence rate in \refE{VVLimit}
    is bounded by $C (\nu t)^{1/2}$.)
\end{proof}

We could have concluded the proof another, more indirect way, as follows. From
\refE{HeatSolution} we see that $\omega(t)$ is continuous in the
$L^2(\Omega)$-norm at time zero. Thus, we can multiply
\refE{VorticityFormulationE} by $\omega$ and integrate over time to give
\begin{align}\label{e:OmegaL2NormConserved}
    \frac{1}{2} \diff{}{t} \norm{\omega(t)}_{L^2(\Omega)}^2
        + \nu \norm{\grad \omega(t)}_{L^2(\Omega)}^2
           &= \nu \int_\Gamma (\grad \omega \cdot \mathbf{n}) \omega.
\end{align}
But $\grad \omega = -(\Delta u)^\perp = 0$ on $\Gamma$ because $\Delta u$ is in
$V$ as we showed, and we conclude by integrating over time that $\omega$ is in
$L^\iny([0, \iny); L^2(\Omega))$, and \refE{VVLimit} follows from the condition
in \refE{OmegaL2Condition} of \refT{KatoKelliherTemamWang}.

This argument shows that whenever the vorticity is continuous at time zero in
the $L^2(\Omega)$-norm and $\Delta u(t)$ lies in $V$ for all $t > 0$,
\refE{VVLimit} holds. The latter condition, however, is very special. In fact,
for $\ol{u}^0$ with regularity as in \refT{VVRadialV}, $\prt_t u$ will always
be in $V$, so $\Delta u = (\prt_t u + u \cdot \grad u + \grad p)/\nu$ is in $V$
if and only if $u \cdot \grad u + \grad p$ is in $V$. This, in turn, can hold
only if $\grad p = -u \cdot \grad u$ on $\Gamma$.

This suggests that for initial velocities for which $\omega(t)$ is continuous
in the $L^2(\Omega)$-norm at time zero we might attempt to make an argument
using \refE{OmegaL2NormConserved}, though now without the right-hand side
vanishing. This would require either control on $\grad \omega$, as we had
above, or on $\omega$ itself. (For solutions to ($NS$) with the boundary
condition $\omega = u \cdot \mathbf{n} = 0$ rather than $u = 0$ this term would
vanish and one can obtain the vanishing viscosity limit easily---though not
using \refT{KatoKelliherTemamWang}, which one would need to show applies to
such solutions---as done, for instance, in \cite{JL1969} and \cite{L1996}.)

Another line of attack is also suggested by the proof of \refT{VVRadialV}---to
gain control on the coefficients $g(t)$ either by assumptions on the initial
velocity or on the solution. Unless bounds as remarkably strong as those
obtained in the proof of \refT{VVRadialV} are achieved, though, something more
sophisticated must be employed to obtain \refE{VVLimit}.

\section{Interpretation in terms of length scales}\label{S:CW2006}

\noindent In \cite{CW2006}, Cheng and Wang consider the vanishing viscosity
limit in the setting of a two-dimensional rectangular channel $R$, periodic in
the $x$ direction with period $L$ and with characteristic boundary conditions
(which include no-slip boundary conditions as a special case). They decompose
any vector $u$ on $R$ of sufficient regularity as $u = \sum_{j = 0}^\iny e^{2
\pi i j x/L} u^j$ and define the projection $P_k u = \sum_{j = 0}^k e^{2 \pi i
j x/L} u^j$ onto the space spanned by the first $k$ modes. This in effect
allows one to isolate successively finer-scale spatial variations in the
direction tangential to the boundary. They then construct an approximation
sequence $\smallset{v^L}$ to $u$ by letting $v^L$ be the solution to the
equation that results after projecting each term in ($NS$) using $P_N$. (We
have changed their notation somewhat.) Their $v^L$ is the approximate-solution
analog of the exact solution truncation represented by $\widetilde{u}^L$ in
\refE{uTildeN}.

The main result in \cite{CW2006} is that $v^{L(\nu)}$ converges to $\ol{u}$ in
$L^\iny([0, T]; L^2(\Omega))$ as $\nu \to 0$. The requirement on $L(\nu)$ is
the same as our condition on $L$ in \refE{LMNCondition} (with the additional
condition that $L(\nu) \to \iny$ as $\nu \to 0$ as one would expect), so
convergence of $v^{L(\nu)}$ to $\ol{u}$ occurs when only tangential length
scales of order larger than $\nu$ are included in the approximations. (All
length scales in the normal direction, however, are included. See
\refR{TryImproveOmega} concerning this issue in regards to the vorticity.)

The result in \cite{CW2006} makes an important observation about the difficulty
of determining numerically whether or not the vanishing viscosity limit holds.
Our method of decomposing the solution using the eigenfunctions of the Stokes
operator, on the other hand, says little about computation, since approximating
this decomposition numerically is probably as least as hard as approximating
the solution itself. Nonetheless, it more directly characterizes the properties
of the solution itself at different length scales.

The analog to the result in \cite{CW2006} is \refT{VanishingutangentialResult},
which shows that Temam and Wang's conditions in \refE{TemamWangCondition1} and
\refE{TemamWangCondition2}, when applied only to the modes with tangential
wavelengths of $C L(\nu)$ or higher, holds as long as the condition on $N$ in
\refE{LMNCondition} hold. This does not, however, imply that $u^{L(\nu)}$
converges to $\ol{u}$ in the vanishing viscosity limit, only that if the
vanishing viscosity limit fails to hold, the failure originates in the behavior
of the tangential component of the gradient projected into the space spanned by
the modes with tangential frequencies of order $L(\nu)$ or higher; that is, at
length scales of order $\nu$ or lower.

The other conditions in \refT{MainResult} give alternative ways to measure the
behavior of the solution at different length scales or frequencies. They show
that we cannot simply say that if the vanishing viscosity fails to hold then
the failure lies in the behavior of the solution at any particular range of
length scales, but rather that the pertinent range of length scales varies with
the measure of behavior. Whether any of these conditions brings us any closer
to proving that the vanishing viscosity limit holds in general for smooth
initial data in a bounded domain or to proving that it fails to hold in at
least one instance remains completely unclear.

\appendix

\section{Bounds on the Eigenfunctions}\label{S:EigenfunctionBounds}

\noindent In \refL{BasicBessel} we state the basic identities involving the
Bessel functions that we use. We then give a series of lemmas that lead to the
bounds on the velocity and vorticity of the eigenfunctions in the boundary
layer that we used in the proof of \refT{MainResult}.

It is perhaps important to note that in the proofs that follow we avoid the use
of asymptotic formulas for the Bessel functions, even when such formulas might
appear to be useful. This is because we need to deal with the relative values
of Bessel functions of different orders near a zero of one of the Bessel
functions, and it is precisely in these situations that the errors in the
asymptotic formulas dominate. Also, most of the following lemmas apply without
change to their proofs with $n$ being any nonnegative real value.

\begin{lemma}\label{L:BasicBessel}
    For all nonnegative real numbers $n$ and $x$,
    \begin{align}\label{e:Bowman628}
        2n J_n(x) - x J_{n -1}(x) = x J_{n + 1}(x),
    \end{align}
    \begin{align}\label{e:Bowman629}
        2 J_n'(x) = J_{n - 1}(x) - J_{n + 1}(x),
    \end{align}
    \begin{align}\label{e:Bowman630}
        J_{n - 1}(x) = \frac{n}{x} J_n(x) + J_n'(x),
    \end{align}
    \begin{align}\label{e:Bowman631}
        J_{n + 1}(x) = \frac{n}{x} J_n(x) - J_n'(x),
    \end{align}
    \begin{align}\label{e:Bowman638}
        \frac{x^n J_n(\al x)}{\al}
           &= \int x^n J_{n - 1} (\al x) \, dx,
    \end{align}
    \begin{align}\label{e:Bowman639}
        J_n(\al x) x^{-n}
            = - \al \int J_{n + 1} (\al x) x^{-n} \, dx,
    \end{align}
    \begin{align}\label{Bowman651}
        (\beta^2 - \al^2) \int x J_n(\al x) J_n(\beta x) \, dx
            = x \brac{\al J_n'(\al x) J_n(\beta x)
                    - \beta J_n'(\beta x) J_n(\al x)},
    \end{align}
    \begin{align}\label{e:Bowman652}
        \int x J_n(a x)^2 \, dx
        &= \frac{1}{2} \brac{x^2 J_n'(a x)^2
                    + \pr{x^2 - \frac{n^2}{a^2}} J_n(a x)^2},
    \end{align}
    \begin{align}\label{e:Bowman653}
        \int x J_n(a x)^2 \, dx
        &= \frac{x^2}{2} \brac{J_n(a x)^2
                    - J_{n - 1}(a x) J_{n + 1}(a x)}.
    \end{align}
\end{lemma}
\begin{proof}
    These are standard identities for Bessel functions. For instance, see
    Equations (6.28), (6.29), (6.30), (6.31), (6.38), (6.39), (6.51), (6.52), and (6.53) of
    \cite{Bowman}.
\end{proof}

\begin{lemma}\label{L:ZeroDifference}
    For all nonnegative integers $n$ and all positive integers $k$,
    \begin{align*}
        1 < j_{n + 1, k} - j_{nk} < \frac{\pi}{2},
    \end{align*}
    where $j_{nk}$ is defined in \refE{jnk}.
\end{lemma}
\begin{proof}
    Let $j_{\nu k}$ be the $k$-th positive zero of $J_\nu$, where we now allow
    $\nu$ to be a real number in the interval $[0, \iny)$. It is shown in
    \cite{Elbert1977} and \cite{ElbertLaforgia1984} that for all $k \ge 1$, $j_{\nu
    k}$ is strictly concave as a function of $\nu$ and that $d j_{\nu k}/d \nu > 1$
    (see also \cite{LaforgiaMuldoon1984}). Thus, the function $n \mapsto j_{n + 1,
    k} - j_{nk}$ is strictly decreasing as a function of $n$. But by Equation (2.9)
    of \cite{ElbertLaforgia1984}, $j_{n + 1, k} - j_{nk} \to 1$ as $n \to \iny$, so
    $j_{n + 1, k} - j_{nk} > 1$.

    The positive zeros of $J_0$ lie in the intervals $(m \pi + \frac{3}{4} \pi, m
    \pi + \frac{7}{8} \pi)$, $m = 0, 1, \dots$, and the positive zeros of $J_1$ lie
    in the intervals $(m' \pi + \frac{1}{8} \pi, m' \pi + \frac{1}{4} \pi)$, $m' =
    1, 2, \dots$. That the zeros lie in only these intervals is shown in Section
    15.32 p. 489 and Section 15.34 p. 491 of \cite{Watson} using an approach of
    Schafheitlin's. That each of these intervals contains at least one zero is
    shown on p. 104 of \cite{Bowman}. But $j_{1 k} - j_{0 k} > 1$ as we showed
    above so each interval contains precisely one zero. Because the zeros of $J_0$
    and $J_1$ are interleaved (see p. 106 of \cite{Bowman}, for instance) we can
    then conclude that $j_{1 k} - j_{0 k} < \frac{\pi}{2}$. But as we observed
    above, the function $n \mapsto j_{n + 1, k} - j_{nk}$ is strictly decreasing as
    a function of $n$, so $j_{n + 1, k} - j_{nk} < \frac{\pi}{2}$ holds for all
    $n \ge 0$.
\end{proof}

\begin{lemma}\label{L:jnkRange}
    For all $n = 0, 1, \dots$ and $k = 1, 2, \dots$,
    \begin{align*}
        n + k < j_{nk} < \pi (n/2 + k) \le \pi (n + k).
    \end{align*}
\end{lemma}
\begin{proof}
    By \refL{ZeroDifference}, for all $n$ and $j$,
    \begin{align*}
        j_{n k}
        &= j_{0 k} + \sum_{m = 1}^n (j_{mk} - j_{m - 1, k})
            \ge j_{0 k} + n
            > n + k,
    \end{align*}
    because $j_{0 k} > k$ (which follows directly from \refE{BesselDE}; see p.
    485-486 of \cite{Watson}, for instance). By an observation in the proof of
    \refL{ZeroDifference} it follows that $j_{0 k} < \pi k$, and a similar
    argument using the inequality $j_{n + 1, k} - j_{nk} < \frac{\pi}{2}$ from
    \refL{ZeroDifference} gives the upper bound on $j_{nk}$.
\end{proof}

\begin{lemma}\label{L:JRatios}
    Let $\al = j_{n + 1, k}$ and $\beta = j_{nk}$. For $n = 0, 1, 2, \dots$ and
    $k = 1, 2, \dots$,
    \begin{align*}  
        \abs{\frac{J_n(\al x)}
                    {J_n(\al)}} \le 1 \text{ if }
            \frac{\beta}{\al} < x < 1.
    \end{align*}
\end{lemma}
\begin{proof}
    Let $g(x) = J_n(\al x)/\smallabs{J_n(\al)}$. From \refE{Bowman631}, $J'_n(\al)
    = (n/\al) J_n(\al)$, so $J'_n(\al)$ has the same sign as $J_n(\al)$. From this
    we conclude that $\abs{g}$ is increasing in a left-neighborhood $N$ of $1$.

    Between each zero of $J_n$ there is exactly one zero of $J_{n + 1}$ (see p. 106
    of \cite{Bowman}, for instance). Between each zero of $J_n$ there is also
    exactly one zero of $J'_n$, because the maximum values of $J_n$ are all
    positive and the minimum values are all negative (see, for instance, p. 107 of
    \cite{Bowman}) and $J_n'$ has no repeated positive roots (this follows from the
    defining equation \refE{BesselDE}). Thus, the neighborhood $N$ includes all $x$
    such that $\beta < \al x < \al$. Since $\abs{g(1)} = 1$ it follows that
    $\abs{g(x)} \le 1$ for all such $x$.
\end{proof}

\begin{lemma}\label{L:Jnp1Ratios}
    Let $\al = j_{n + 1, k}$ and $\beta = j_{nk}$. There exists a constant $C$ such
    that for all $n = 0, 1, \dots$ and $k = 1, 2 \dots$ n,
    \begin{align*}
        \abs{\frac{J_{n + 1}(\al x)} {J_n(\al)}} \le C n (1 - x)
            \text{ if } \frac{\beta}{\al} < x < 1.
    \end{align*}
\end{lemma}
\begin{proof}
    Since $J_{n + 1}(\al) = 0$, \refE{Bowman638} with $n + 1$ in place of $n$ gives
    \begin{align*}
        J_{n + 1}(\al x)
        &= - \frac{\al}{x^{n + 1}} \int_x^1 t^{n + 1} J_n(\al t) \, dt.
    \end{align*}
    As long as $\beta < \al x < \al$,  $J_n(\al t)$ does not change sign on the
    interval $(x, 1]$ and has its maximum value on this interval at $1$, as
    observed in the proof of \refL{JRatios}. Thus,
    \begin{align*}
        \abs{J_{n + 1}(\al x)}
        &\le \frac{\al}{x^{n + 1}} \abs{J_n(\al)}  \int_x^1 t^{n + 1} \, dt
            \le \frac{\al \abs{J_n(\al)}}{(\beta/\al)^{n + 1}} (1 - x).
    \end{align*}
    But by \refL{ZeroDifference} and \refL{jnkRange},
    \begin{align*}
        1 - \frac{\beta}{\al} = \frac{\al - \beta}{\al} \le \frac{\pi/2}{n + 2}
            \implies \frac{\beta}{\al} \ge 1 - \frac{\pi}{2n + 4}
    \end{align*}
    so
    \begin{align*}
        (\beta/\al)^{-(n + 1)}
        &\le \pr{1 - \frac{\pi}{2n + 4}}^{-(n + 1)}
         \le e^{\pi/2},
    \end{align*}
    the last inequality following from elementary calculus. We
    conclude that
    \begin{align}\label{e:betaalRatioBound}
        \abs{J_{n + 1}(\al x)}
        &\le C n \abs{J_n(\al)}(1 - x),
    \end{align}
    which completes the proof.
\end{proof}

\begin{lemma}\label{L:Jnm1Ratios}
Let $\al = j_{n + 1, k}$ and $\beta = j_{nk}$. There exists a constant $C$ such
that for all $n = 0, 1, \dots$ and $k = 1, 2 \dots, n$,
    \begin{align*}
        \abs{\frac{J_{n - 1}(\al x)} {J_n(\al)}} \le C
            \text{ if } \frac{\beta}{\al} < x < 1.
    \end{align*}
\end{lemma}
\begin{proof}
    Because the positive zeros of $J_{n - 1}$ are interlaced with those of $J_n$,
    $J_{n - 1}$ does not change sign on the interval $[\beta, \al]$. From
    \refE{Bowman631} with $n - 1$ in place of $n$, $J'_{n - 1}(\beta) = ((n -
    1)/\beta) J_{n - 1}(\beta)$, so $J'_{n - 1}(\beta)$ has the same sign as $J_{n
    - 1}(\beta)$, and we conclude that $J_{n - 1}$ reaches it maximum value on the
    interval $[\beta, \al]$ at $\beta$. Therefore, for $\beta/\al < x < 1$,
    \begin{align*}
        \abs{\frac{J_{n - 1}(\al x)} {J_n(\al)}}
            \le \abs{\frac{J_{n - 1}(\beta)} {J_n(\al)}}.
    \end{align*}
    But, by \refE{Bowman628}, $J_{n + 1}(\beta) = 2 (n/\beta) J_n(\beta) - J_{n -
    1}(\beta) = - J_{n - 1}(\beta)$, so
    \begin{align*}
        \abs{\frac{J_{n - 1}(\al x)} {J_n(\al)}}
            \le \abs{\frac{J_{n + 1}(\beta)} {J_n(\al)}}
            \le C n (1 - \beta/\al)
            \le C,
    \end{align*}
    where we used \refL{Jnp1Ratios} and \refL{jnkRange}.
\end{proof}

\begin{lemma}\label{L:L2omegaGammaBound}
    We have $\norm{\omega_j}_{L^2(\Gamma_{\delta})}^2 \le 2
    \delta$ when $\delta \le \la_j^{-1/2}$.
\end{lemma}
\begin{proof}
    Let $\omega_j = \omega_{nk}$ and $\al = j_{n + 1, k} = \la_j^{1/2}$. Then
    \begin{align*}
        \norm{\omega_j}_{L^2(\Gamma_{\delta})}^2
        &= 2 \pi C_{nk}^2 \int_{1 - \delta}^1
                r J_n(\al r)^2 \, dr
            = 2 \int_{1 - \delta}^1 r \frac{J_n(\al r)^2}{J_n(\al)^2}  \, dr.
    \end{align*}
    In the integrals above, with $\beta = j_{nk}$,
    \begin{align*}
        \beta/\al
            = 1 - (\beta - \al)/\al
            \le 1 - 1/\al
            = 1 - \la_j^{-1/2}
            \le 1 - \delta
            < r
            < 1,
    \end{align*}
    where we used \refL{ZeroDifference}, and the lemma follows from
    \refL{JRatios}.
\end{proof}

Employing \refL{L2omegaGammaBound}, we can extend its range of applicability,
though with a higher bound on the width of the boundary layer.

\begin{lemma}\label{L:L2omegaGammaBoundGeneral}
    For all $n = 0, 1, \dots$, $k = 1, \dots, n$, and all $\delta < 2 \pi
    \la_{n1}^{-1/2}$,
    \begin{align*}
        \norm{\omega_{nk}}_{L^2(\Gamma_{\delta})}^2
            \le 2 \delta.
    \end{align*}
\end{lemma}
\begin{proof}
    It follows from \refL{jnkRange} that $j_{n, n}/j_{n, 1} \le \pi (n + n)/(n + 1) \le 2
    \pi$; the lemma follows from this inequality and \refL{L2omegaGammaBound}.
\end{proof}

\begin{remark}\label{R:L2omegaGammaBoundGeneral}
    It is possible to extend \refL{L2omegaGammaBoundGeneral} to include all values
    of $k$. The idea of the proof is that for $k > n$, $\omega_{nk}$ passes through
    $k$ complete half-periods (annuli in the unit disk lying between successive
    nonnegative zeroes of $J_n(j_{n + 1, k} r)$) and ends with a partial period.
    Since $J_n(x)$ decays like $x^{1/2}$ and the spacing between consecutive zeros
    of $J_n$ approaches a constant, the $L^2$-norms of $\omega_{nk}$ on each of
    those half-periods converges to a constant, and since the $L^2$-norm of
    $\omega_{nk}$ on the entire unit disk is 1, the square of the $L^2$-norm of
    $\omega_{nk}$ on the last half-period is less than $C/k$ (with $C$ near 1).
    But the last half-period has a width that is greater than $C/k$.
    Extending this argument to $m$ periods, what we have shown is that
    \begin{align*}
        \norm{\omega_{nk}}_{L^2(\Gamma_{Cm/k})}^2 \le m/k.
    \end{align*}
    With the assumed bound on $\delta$, we choose $m$ so that $m/k$ is of the same
    order as $\delta$, and the proof is essentially complete.
\end{remark}

\begin{lemma}\label{L:L2uGammaBoundGeneral}
    There exist positive constants $C_1$ and $C_2$ with $C_2 < 1$ such that for
    all $n = 0, 1, \dots$ and all $k = 1, \dots, n$,
    \begin{align*}
        \norm{u_{nk}}_{L^2(\Gamma_{\delta})}^2
            \le C_1 \delta^3
    \end{align*}
    when $\delta < C_2 \la_{n1}^{-1/2}$.
\end{lemma}
\begin{proof}
    In the proof that follows, we will often use \refL{jnkRange} without explicit
    mention.

    Let $\al = j_{n + 1, k}$. We bound first the radial component of $u_{nk}$.
    We have,
    \begin{align*}
        \frac{J_n(\al r) - J_n(\al) r^n}{J_n(\al) r}
            = r^{n - 1} g_n(r),
    \end{align*}
    where
    \begin{align*}
        g_n(r)
        &=\frac{J_n(\al r) r^{-n}}{J_n(\al)} - 1
            = -\frac{\al}{J_n(\al)} \int_r^1 \frac{J_{n + 1}(\al x)}{x^n} \, dx
        &= -\al \int_r^1 \frac{B_{nk}(x)}{x^n} \, dx,
    \end{align*}
    and
    \begin{align*}
        B_{nk}(x) = \frac{J_{n + 1}(\al x)}{J_n(\al)}.
    \end{align*}
    To verify the second equality we use the identity in \refE{Bowman639}, from
    which it follows that the second expression for $g_n$ is
    \begin{align*}
        \frac{J_n(\al r) r^{-n} + C}{J_n(\al)}
    \end{align*}
    for some constant $C$. But all three expressions for $g_n$ are zero at $r = 1$,
    so we have the correct limits of integration in the second expression. It
    follows from \refL{Jnp1Ratios} and our third expression for $g_n$ that
    \begin{align*}
        \abs{g_n(r)}
            \le C n \al \int_r^1 x^{-n} (1 - x) \, dx
            \le \frac{C n^2}{r^n} (1 - r)^2
    \end{align*}
    for all $1 - r \le 1/\al$.

    From \refE{unk},
    \begin{align*}
        u_{nk}^r(r, \theta)
        &= \frac{g(r) r^{n - 1}}
                {\al^2 \pi^{1/2}}
                        i n e^{i n \theta} \wh{e}_r,
    \end{align*}
    so when $\delta \le 1/\al$ we have
    \begin{align*}
        \norm{u_{nk}^r}_{L^2(\Gamma_\delta)}^2
        &= 2 \pi \int_{1 - \delta}^1 r \abs{u_{nk}^r}^2 \, dr
            \le \frac{C n^6}{\al^4}
                \int_{1 - \delta}^1 \frac{(1 - r)^4}{r^{2n - 1}} \, dr \\
        &\le C n^2 (1 - \delta)^{1 - 2n} \delta^5
            \le C \delta^3.
    \end{align*}
    In the last inequality we used
    \begin{align*}
        \delta
           &< C_2 \la_{n1}^{-1/2}
            = \frac{C_2}{j_{n + 1, k}}
            \le \frac{C_2}{n + 1}
    \end{align*}
    so
    \begin{align*}
        (1 - \delta)^{2n - 1}
        &\ge \pr{1 - \frac{C_2}{n + 1}}^{2n - 1}
         = \pr{G(C_2, n + 1)}^2 \pr{1 - \frac{C_2}{n + 1}}^{-3} \\
        &\ge (1 - C_2)^2 \pr{1 - C_2}^{-3}
         = (1 - C_2)^{-1}
         = C
         > 0,
    \end{align*}
    where $G$ is the function of \refL{UsefulFunctionBound}.

    For the angular component of $u_{nk}$, we write
    \begin{align*}
        \al &J_{n + 1}(\al r) - \al J_{n - 1}(\al r) + 2n J_n(\al) r^{n - 1} \\
        &= \brac{2n J_n(\al r) - \al r J_{n - 1}(\al r)}
                + \al r J_{n - 1}(\al r) - 2n J_n(\al r) \\
            &\qquad\qquad
                + \al J_{n + 1}(\al r) - \al J_{n - 1}(\al r)
                + 2n J_n(\al) r^{n - 1} \\
        &= \al(r + 1) J_{n + 1}(\al r)
                + 2n \brac{J_n(\al) r^{n - 1} - J_n(\al r)} \\
            &\qquad\qquad
                + \al J_{n - 1}(\al r)(r - 1).
    \end{align*}
    From \refE{unk} we then have
    \begin{align}\label{e:InitunkThetaBound}
        \begin{split}
            \smallabs{u_{nk}^\theta}^2
            &\le C \frac{\al^2 (r + 1)^2}{\al^4} \frac{J_{n + 1}(\al r)^2}{J_n(\al)^2}
                + C \frac{n^2}{\al^4} \pr{\frac{J_n(\al) r^{n - 1} - J_n(\al
                    r)}{J_n(\al)}}^2 \\
            &\qquad + C \frac{\al^2 J_{n - 1}(\al r)^2 (1 - r)^2}{\al^4 J_n(\al)^2}
                \\
            &\le C (1 - r)^2
                + \frac{C}{n^2} \pr{r^{n - 2} g_{n - 1}(r)}^2,
        \end{split}
    \end{align}
    where we applied both \refL{Jnp1Ratios} and \refL{Jnm1Ratios}.

    The first term in \refE{InitunkThetaBound} contributes no more
    \begin{align*}
        C \int_{1 - \delta}^1
                    r (1 - r)^2 \, dr
            \le C \int_{1 - \delta}^1
                    (1 - r)^2 \, dr
            \le C \delta^3,
    \end{align*}
    and the same is true of the second term in \refE{InitunkThetaBound} arguing as
    for $u_{nk}^r$, and this completes the proof.
\end{proof}

\begin{lemma}\label{L:SomeL2InnerProductsAreZero}
    When $m \ne n$, $\innp{u_{mj}, u_{nk}}_{L^2(\Gamma_{c \nu})} =
    \innp{\omega_{mj}, \omega_{nk}}_{L^2(\Gamma_{c \nu})} = 0$.
\end{lemma}
\begin{proof}
    We have,
    \begin{align*}
    \innp{\omega_{mj}, \omega_{nk}}_{L^2(\Gamma_{c \nu})}
        = \int_{1 - c \nu}^1 r f(r) \int_0^{2 \pi} e^{i (m - n) \theta}
                \, d \theta \, dr,
    \end{align*}
    where $f(r)$ is a product of two Bessel functions. When $m \ne n$, the
    inner integral is zero. A similar argument gives $\innp{u_{mj}, u_{nk}}_{L^2(\Gamma_{c \nu})}
    = 0$.
\end{proof}

\begin{lemma}\label{L:UsefulFunctionBound}
    Let $\al$ be in $(0, 1)$ and define $G_\alpha: [1, \iny) \to [0, \iny)$ by
    \begin{align*}
        G_\alpha(x) = \pr{1 - \frac{\al}{x}}^x.
    \end{align*}
    Then for all $x > 1$,
    \begin{align*}
        1 - \al \le G_\alpha(x) < e^{-\al}.
    \end{align*}
\end{lemma}
\begin{proof}
    The proof is elementary.
\end{proof}

\Tentative{

    \newpage

    \begin{lemma}\label{L:L2graduGammaBound}
        We have $\norm{\grad u_j}_{L^2(\Gamma_{\delta})}^2 \le 2
        \delta$ when $\delta \le \la_j^{-1/2}$.
    \end{lemma}
    \begin{proof}
    Let $u_j = u_{nk}$ and $\al = j_{n + 1, k} = \la_j^{1/2}$. Then
    \begin{align*}
        \norm{\grad u_j}_{L^2(\Gamma_\delta)}^2
        &= \int_{\Gamma_\delta} \grad u_{nk} \cdot \ol{\grad u_{nk}} \\
        &= - \int_{\Gamma_\delta} \Delta u_{nk} \cdot \ol{u_{nk}}
                    + \int_{\prt \Gamma_\delta} (\grad u_{nk} \cdot \mathbf{n})
                                \cdot \ol{u_{nk}}.
    \end{align*}
    But $\Delta u_{nk} + \la_{nk} u_{nk} = \grad p_{nk}$ for some scalar field
    $p_{nk}$ by \refE{StokesEigenfunctionSolution} so
    \begin{align*}
        -\int_{\Gamma_\delta} &\Delta u_{nk} \cdot \ol{u_{nk}}
            = \la_{nk} \int_{\Gamma_\delta} u_{nk} \cdot \ol{u_{nk}}
                + \int_{\Gamma_\delta} \grad p_{nk} \cdot \ol{u_{nk}} \\
        &= \al \norm{u_{nk}}_{L^2(\Gamma_{\delta})}^2
                - \int_{\prt \Gamma_\delta} (\ol{u_{nk}} \cdot \mathbf{n}) p_{nk}.
    \end{align*}

    Now, $u_{nk}$ is of the form $(F^r(r) \wh{e}_r + F^\theta(r) \wh{e}_\theta)
    e^{i n \theta}$ and it is clear without even calculating what $p_{nk}$ is that
    $p_{nk} = G(r) e^{i n \theta}$, else \refE{StokesEigenfunctionSolution} could
    not hold. Then since $u_{nk} = 0$ on the outer boundary of $\Gamma_\delta$
    (namely, $\Gamma$), and $\mathbf{n} = - \wh{e}_r$ on the inner boundary of
    $\Gamma_\delta$, we have
    \begin{align*}
        \int_{\prt \Gamma_\delta} (\ol{u_{nk}} \cdot \mathbf{n}) p_{nk}
        &= - 2 \pi (1 - \delta) G(1 - \delta) F^r(1 - \delta)
    \end{align*}

    \textbf{Fine}, but at this point we need to use the formula for $p_{nk}$:
    \begin{align*}
        p_{0k}(r, \theta) = 0, \;
        p_{nk}(r, \theta) = C r^n e^{i n \theta},
    \end{align*}
    being, I believe, correct, though one would have to determine the constants.

    ....

    This leaves
    \begin{align*}
        \int_{\prt \Gamma_\delta} (\grad u_{nk} \cdot \mathbf{n})
                                \cdot \ol{u_{nk}}
            = \frac{1}{2} \int_{\prt^1 \Gamma_\delta} \grad \abs{u_{nk}}^2 \cdot
            \mathbf{n},
    \end{align*}
    I believe, where $\prt^1 \Gamma_\delta$ is the inner boundary of
    $\Gamma_\delta$. Then
    \begin{align*}
        \frac{1}{2} \int_{\prt^1 \Gamma_\delta} &\grad \abs{u_{nk}}^2 \cdot \mathbf{n}
            = \frac{1}{2} \int_{\prt^1 \Gamma_\delta}
                \pr{\pdx{\abs{u_{nk}}^2}{r} \wh{e}_r
                    + \pdx{\abs{u_{nk}}^2}{\theta} \wh{e}_\theta}
                        \cdot (- \wh{r}) \\
        &= - \frac{1}{2} \int_{\prt^1 \Gamma_\delta} \pdx{\abs{u_{nk}}^2}{r}
            = - \pi(1 - \delta) \pdx{\abs{u_{nk}}^2}{r}\big|_{r = 1 - \delta}.
    \end{align*}

    Now,
    \begin{align*}
        \pdx{\abs{u_{nk}}^2}{r}
        &= 2 u_{nk} \cdot \pdx{\ol{u_{nk}}}{r}
    \end{align*}

    ...............
    \end{proof}
    }

\Obsolete{
    \begin{lemma}\label{L:Jnalphabound}
        There exists a positive constant $C$ such that
        \begin{align*}
            J_n(j_{n + 1, k}) \ge C j_{n + 1, k}^{-1/2}
        \end{align*}
        for all $n = 0, 1, \dots$ and $k = 1, 2, \dots$.
    \end{lemma}
    \begin{proof}
        What we must show is that the function $x^{1/2} J_n(x) \ge C$ for some constant
        $C$ when evaluated at any zero of $J_n(x)$. Let $\al = j_{n + 1, k}$, and
        observe that \refE{Bowman630} applied with $n + 1$ in place of $n$ gives
        $J_n(\al) = J_{n + 1}'(\al)$, so we have $\al^{1/2} J_n(\al) = \al^{1/2} J_{n +
        1}'(x)$. It will suffice to show, then, that the function $P(x) = x^{1/2} J_{n
        + 1}'(x) \ge C$ when evaluated at $\al$.

        Letting $u(x) = x^{1/2} J_{n + 1}(x)$, $u$ satisfies the differential equation,
        \begin{align}\label{e:Jn1ODE}
            u''(x) = f(x) u, \text{ where }
                f(x) = - \pr{1 - \frac{(n + 1)^2 - 1/4}{x^2}}.
        \end{align}
        (See, for instance, Equation (6.57) of \cite{Bowman}.) Then
        \begin{align}\label{e:up}
            u'(x)
                = \frac{1}{2} x^{-1/2} J_n(x) + \sqrt{x} J_{n + 1}'(x)
                = \frac{1}{2} x^{-1/2} J_n(x) + P(x),
        \end{align}
        and we have by asymptotic formulas\MarginNote{This might have a problem for $k
        = 1$.}for $J_n(x)$ that $\abs{J_n(x)} \le C x^{-1/2}$, so the first term is of
        order $1/x$. This will allow us to use $u'(x)$ as a good enough estimate of
        $P(x)$, as we shall see.

        Multiplying the differential equation in \refE{Jn1ODE} by $u'$ gives
        \begin{align*}
            u'' u' = &f(x) u u'
                \implies \frac{1}{2} \frac{d}{dx} (u')^2
                    = \frac{1}{2} f(x) \frac{d}{dx} u^2 \\
            &\implies \frac{d}{dx} (u')^2
                    = \frac{d}{dx} (f(x) u^2) - f'(x) u^2.
        \end{align*}
        But,
        \begin{align*}
            f'(x) = -2 \frac{(n + 1)^2 - 1/4}{x^3},
        \end{align*}
        and it follows that
        \begin{align*}
            \frac{d}{dx} (u')^2
                    = \frac{d}{dx} (f(x) u^2) + g(x),
        \end{align*}
        where $g(x) = - f'(x) u(x)^2 \ge 0$. Integrating this equation from $\gamma =
        j_{n + 1, k - 1}$  to $\al$ (assuming that $k \ge 2$) gives
        \begin{align}\label{e:upDiff}
            \begin{split}
                u'(\al)^2 - u'(\gamma)^2
                &= f(\al) u(\al)^2 - f(\gamma) u(\gamma)^2
                        + \int_\gamma^\al g(x) \, dx \\
                &= \int_\gamma^\al g(x) \, dx
                        > 0,
            \end{split}
        \end{align}
        and we conclude that $u'(\al)$ increases with the value of $k$. Since the
        difference between $P(\al)$ and $u'(\al)$ is of order $1/x$, as noted above,
        and the first positive zero of $J_{n + 1}$ is greater than $n$, we can see that
        the problem is reduced to the case $k = 1$.

        So now assume that $k = 1$ and that $n \ge 1$, and let $\gamma = 0$. Then $J_{n
        + 1}(\gamma) = J_{n + 1}'(\gamma) = 0$, so $u(\gamma) = 0$ and also, from
        \refE{up}, $u'(\gamma) = 0$. Then from \refE{upDiff},
        \begin{align*}
            u'(\al)^2
                &= \int_\gamma^\al g(x) \, dx
                \ge 2 \frac{(n + 1)^2 - 1/4}{\al^3}
                \int_0^\al u(x)^2 \, dx.
        \end{align*}
        But by \refE{Bowman652},\MarginNote{This reaches a dead end!}
        \begin{align*}
            \int_0^\al u(x)^2 \, dx
            &= \int_0^\al x J_{n + 1}(x)^2 \, dx \\
            &= \frac{1}{2} \brac{x^2 J_{n + 1}'(x)^2
                            + \pr{x^2 - (n + 1)^2} J_{n + 1}(x)^2}_0^\al \\
            &= \frac{\al^2}{2} J_{n + 1}'(\al)^2
        \end{align*}
    \end{proof}
    }

%
%
\Obsolete{
    \begin{lemma}\label{L:L2omegaGammaBoundGeneral}
        There exist $C, C' > 0$ such that for all $j = 1, 2, \dots$,
        \begin{align*}
            \norm{\omega_{nj}}_{L^2(\Gamma_{\delta})}^2
                \le C \delta
        \end{align*}
        when $\delta < C' \la_{n1}^{-1/2}$.
    \end{lemma}
    \begin{proof}
        It\MarginNote{We only need \refL{L2omegaGammaBoundGeneral} for $j$ up to $n$,
        which means we should remove the remainder of the proof, which is kind of
        casual anyway, and replace it with a comment after the proof of the
        lemma.}follows from asymptotic expressions for Bessel functions that
        \begin{align*} 
            j_{n, n}/j_{n, 1} \le C.
        \end{align*}
        The lemma follows for $j \le n$ from this inequality and
        \refL{L2omegaGammaBound}.

    For $j > n$, $\omega_{nj}$ passes through $j$ complete half-periods (annuli in
    the unit disk lying between successive nonnegative zeroes of $J_n(j_{n + 1, k}
    r)$) and ends with a partial period. Since\MarginNote{Obviously need to be more
    precise here.}$J_n(x)$ decays like $x^{1/2}$ and the spacing between
    consecutive zeros of $J_n$ approaches a constant, the $L^2$-norms of
    $\omega_{nj}$ on each of those half-periods converges to a constant, and since
    the $L^2$-norm of $\omega_{nj}$ on the entire unit disk is 1, the square of the
    $L^2$-norm of $\omega_{nj}$ on the last half-period is less than $C/j$ (the
    $L^2$-norm squared on the disk is the sum of the $L^2$-norm squared on
    partitions of the disk). But the last half-period has a width that is
    greater\MarginNote{$C$ is about 0.7.}than $C/j$. In other words, what we have
    shown is that
    \begin{align*}
        \norm{\omega_{nj}}_{L^2(\Gamma_{c_0/j})}^2 \le 1/j
    \end{align*}
    for some positive constant $c_0$.

    As $r$ increases, the half-periods decrease in width, so extending the
    reasoning above to the last $k$ half-periods it follows that
    \begin{align*}
        \norm{\omega_{nj}}_{L^2(\Gamma_{c_0k/j})}^2 \le k/j.
    \end{align*}
    Let $k = [\delta j/c_0] + 1$. Then $\delta \le c_0k/j$ so
    \begin{align}\label{e:IntOmegaGammaBound}
        \norm{\omega_{nj}}_{L^2(\Gamma_\delta)}^2
        &\le k/j
            \le \frac{\frac{\delta j}{c_0} + 1}{j}
            = \frac{\delta}{c_0} + \frac{1}{j}.
    \end{align}

    Now, by \refL{ZeroDifference}, for all $n$ and $j$,
    \begin{align*}
        j_{n, k}
        &= j_{0, k} + \sum_{m = 1}^n (j_{mk} - j_{m - 1, k})
            \le j_{0, k} + Cn
            \le c_1(n + k),
    \end{align*}
    because $j_{0, k} \le C k$ by asymptotic formulas for $J_0$.

    If $\delta > c_1/(2j)$ then the right-hand side of \refE{IntOmegaGammaBound} is
    less than $C \delta$ and the lemma holds. If $\delta \le c_1/(2j)$ then
    \begin{align*}
        \delta
        &\le \frac{c_1}{2j}
            \le \frac{c_1}{n + 1 + j}
            \le \frac{1}{j_{n + 1, j}}
            = \la_{nj}^{-1/2}
    \end{align*}
    and the lemma holds by directly applying \refL{L2omegaGammaBound}. Since this
    covers all the cases, the proof is complete.
    \end{proof}
    }

%
%
\Obsolete{
    \begin{lemma}\label{L:Spec2}
        There\MarginNote{We are not using this lemma anymore, so remove it.}exists a
        $C > 0$ such that for all $n = 0, 1, \dots$ and $j, k = 1, 2, \dots$, $j \ne k$,
        and all $\delta$ in $[0, 1]$,
        \begin{align*}
            \smallabs{\innp{\omega_{nj}, \omega_{nk}}_{L^2(\Gamma_{\delta})}}
                \le \frac{C}{\abs{k - j}}.
        \end{align*}
    \end{lemma}
    \begin{proof}
    Let $\al = j_{n + 1, j}$ and $\beta = j_{n + 1, k}$. Define a function $I: [0,
    1] \to \R$ by
    \begin{align*}
        I(x)
            &:= \innp{\omega_{nj}, \omega_{nk}}_{L^2(\Gamma_{1 - x})}
            = 2 \pi C_{nj} C_{nk}
                \int_x^1 r J_n(\al r) J_n(\beta r) \, dr \\
            &= \frac{2}{J_n(\al) J_n(\beta)}
                \int_x^1 r J_n(\al r) J_n(\beta r) \, dr.
    \end{align*}
    Using the identity in \refE{Bowman651} gives
    \begin{align*}
        I(x)
        &= \frac{2}{J_n(\al) J_n(\beta) (\beta^2 - \al^2)}
                \brac{r \al J_n'(\al r) J_n(\beta r)
                    - r \beta J_n'(\beta r) J_n(\al r)}_x^1.
    \end{align*}
    But as in \refE{EigenRoot}, $J_n'(x) = (n/x) J_n(x) - J_{n + 1}(x)$, so
    \begin{align*}
        J_n'(\al) = \frac{n}{\al} J_n(\al) \text{ and }
        J_n'(\beta) = \frac{n}{\beta} J_n(\beta)
    \end{align*}
    and the upper limit above becomes
    \begin{align*}
        \al J_n'(\al) J_n(\beta) - \beta J_n'(\beta) J_n(\al)
            = 0
    \end{align*}
    so
    \begin{align*}
        I(x)
        &= \frac{2 x (\al J_n'(\al x) J_n(\beta x)
                    - \beta J_n'(\beta x) J_n(\al x))}
                {J_n(\al) J_n(\beta) (\beta^2 - \al^2)}.
    \end{align*}

    At a maximum of $\abs{I(x)}$ we have $I'(x) = 0$. This derivative is easily
    calculated using the integral expression for $I(x)$, and we conclude that
    \begin{align*}
        y J_n(\al y) J_n(\beta y) = 0
    \end{align*}
    when $\abs{I(y)}$ is a maximum of $\abs{I(x)}$. And thus either $y = 0$ (where
    the value of $I(y)$ is zero) $J_n(\al y) = 0$ or $J_n(\beta y) = 0$.

    Assume that $J_n(\beta y) = 0$. Then
    \begin{align*}
        J_n'(\beta y)
        &= \frac{n}{\beta y} J_n(\beta y) - J_{n + 1}(\beta y)
            = - J_{n + 1}(\beta y)
    \end{align*}
    and
    \begin{align*}
        \abs{I(x)}
        &\le \abs{\frac{2 y \beta J_n'(\beta y) J_n(\al y)}
                {J_n(\al) J_n(\beta) (\beta^2 - \al^2)}} \\
        &\le \frac{2}{\abs{J_n(\al)}\abs{J_n(\beta)}}
                \frac{1}{\abs{\al - \beta}}
                \frac{\beta}{\al + \beta}
                \smallabs{y^{1/2} J_{n + 1}(\beta y)}
                \smallabs{y^{1/2} J_n(\al y)} \\
        &\le \frac{2}{\abs{J_n(\al)}\abs{J_n(\beta)}
                    \abs{\al - \beta}}
                \frac{\abs{f_{n + 1}(\beta y)}}{\beta^{1/2}}
                \frac{\abs{f_n(\al y)}}{\al^{1/2}}
    \end{align*}
    where
    \begin{align*}
        f_n(x) = x^{1/2} J_n(x),
    \end{align*}
    which is bounded by a constant $C$ independent of $n$. Also, $\al^{1/2}
    \abs{J_n(\al)} \ge C$ with $C$ independent of $n$ (this is a somewhat involved
    thing to show, since it is not true for all $x$) and similarly for $\beta$. We
    conclude that
    \begin{align*}
        \abs{I(x)}
        &\le \frac{C}{\al^{1/2} \abs{J_n(\al)} \beta^{1/2} \abs{J_n(\beta)}
                \abs{\al - \beta}} \\
        &\le \frac{C}{\abs{\al - \beta}}.
    \end{align*}
    \end{proof}
    }

\Obsolete{
    \begin{lemma}\label{L:EigenSizes}
        There\MarginNote{Remove this lemma if we end up not using it.}exist constants
        $C_1 > 0$ and $C_2 > C_1$ such that $C_1 p \le \la_p \le C_2 p$ for all $p$. If
        we let $c_{pq} = 0$ when $\innp{\omega_p, \omega_q}_{L^2(\Gamma_{c \nu})} = 0$
        and $c_{pq} = 1$ otherwise, then
        \begin{align*}
            \sum_{p = 1}^{N} \sum_{q = 1}^{N} c_{pq}
                \le C N^{3/2}.
        \end{align*}
    \end{lemma}
    \begin{proof}
        The first property, that $C_1 p \le \la_p \le C_2 p$ for all $p$, holds for a
        general bounded domain in $\R^2$. We give a proof that is specific to the unit
        disk, however, because we need an estimate developed in its proof to establish
        the second property.

        For any $x > 0$, let $J(n, x)$ be the total number of positive zeros of
        $J_n$ that are less-than-or-equal-to $x$. By \refL{ZeroDifference} and
        because $j_{0, k} \ge k$,
        \begin{align*}
            j_{n, k}
            &= j_{0, k} + \sum_{m = 1}^n (j_{mk} - j_{m - 1, k})
            \ge j_{0, k} + n
            \ge n + k.
        \end{align*}
        Therefore, $j_{nk} \le x \implies n + k \le x \implies k \le x - n \implies
        J(n, x) \le x - n$. Hence $J(n, x) = 0$ when $n \ge x$ and the total number of
        positive zeros of Bessel functions $J_n$, $n = 0, 1, 2, \dots$ is bounded above
        by
        \begin{align*}
            \sum_{n = 0}^{[x]} (x - n)
                &= (x)[x] - [x]([x] - 1)/2
                \le x^2.
        \end{align*}
        This means that the total number of eigenvalues $\la_p$ for which $\la_p^{1/2}
        \le x$ is bounded above by $x^2$ and so the total number of eigenvalues for
        which $\la_p \le x$ is bounded above by $x$. Similarly, the total number of
        eigenvalues for which $\la_p \le x$ is bounded below by $C x$, $C < 1$. Thus,
        the growth of the eigenvalues is controlled above and below by linear
        functions, from which it follows that $C_1 p \le \la_p \le C_2 p$.

        But \refL{SomeL2InnerProductsAreZero} means that
        \begin{align*}
            \sum_{p = 1}^{N} \sum_{q = 1}^{N} c_{pq}
                &= \sum_{n = 1}^\iny J(n, \la_N^{1/2})^2
                \le \sum_{n = 1}^{[\la_N^{1/2}]} (\la_N^{1/2} - n)^2
                \le \sum_{n = 1}^M (M - n)^2 \\
                &= \sum_{m = 1}^{M - 1} m^2
                = \frac{M^3}{3} - \frac{M^2}{2} + \frac{M}{6},
        \end{align*}
        where $M = [\la_N^{1/2}] + 1$. But
        \begin{align*}
            \frac{M^3}{3} - \frac{M^2}{2} + \frac{M}{6}
                \le C M^3
                \le C \la_N^{3/2}
                \le C N^{3/2},
        \end{align*}
        which completes the proof.
    \end{proof}
    }

\Obsolete{
    \begin{lemma}\label{L:JProperties}
        There exist constant $C_1 > 0$ and $C_2 > C_1$ such that $C_1 p \le \la_p \le C_2
        p$ for all $p$. Define a function $J: \Z^+ \cup \set{0} \times \Z^+ \to \Z^+$
        by $\la_{mj} \le \la_p \iff j \le J(m, p)$. Then $J(n, x) \le C x$ for all $n$,
        $x$, and for all $p$, $J(m, p) = 0$ for all but a finite number of values of
        $m$.
    \end{lemma}
    \begin{proof}
        By \refL{ZeroDifference},
        \begin{align*}
            j_{n, k}
            &= j_{0, k} + \sum_{m = 1}^n (j_{mk} - j_{m - 1, k})
            \le j_{0, k} + Cn
            \le C(n + k),
        \end{align*}
        because $j_{0, k} \le C k$ by asymptotic formulas for $J_0$. Therefore,
        given $x > 0$, there can be at most $x/C - n$ positive zeros of $J_n$ less
        than $x$; that is, $k < x/C - n$. Hence there are no positive zeros when $n
        > x/C$ and the total number of positive zeros of Bessel functions $J_n$, $n
        = 0, 1, 2, \dots$ is bounded above by
        \begin{align*}
            \sum_{n = 0}^{[x/C]} (x/C - n)
            &= (x/C)[x/C] - [x/C]([x/C] - 1)/2
                \le C x^2.
        \end{align*}
    This means that the total number of eigenvalues $\la_p$ for which $\la_p^{1/2}
    \le x$ is bounded above by $C x^2$ and so the total number of eigenvalues for
    which $\la_p \le x$ is bounded above by $C x$. Similarly, the total number of
    eigenvalues for which $\la_p \le x$ is bounded below by $C' x$, $C' < C$. Thus,
    the growth of the eigenvalues is controlled above and below by linear
    functions, from which it follows that $C_1 p \le \la_p \le C_2 p$.

    But then,
    \begin{align*}
        \la_{mj} \le \la_p
            \iff C(m + j) \le C' p
            \iff j \le C(p - m),
    \end{align*}
    so $J(m, p) \le C(p - m)$. From this, the theorem's remaining two conclusions
    follow.
    \end{proof}
    }

\bibliography{Refs}
\bibliographystyle{plain}

\end{document}